\documentclass[a4paper,11pt]{article}
\usepackage{jheppub} 
\usepackage{lineno}

\arxivnumber{2304.10865} 
\usepackage{lipsum}
\usepackage{blindtext}
\usepackage{graphicx}

\usepackage{dcolumn}
\usepackage{bm}

\usepackage{hyperref}
\usepackage{slashed}
\usepackage{amsfonts}
\usepackage{mathrsfs}
\usepackage{color}
\usepackage[makeroom]{cancel}
\usepackage{graphicx}
\usepackage{amssymb}
\usepackage{amsmath}
\usepackage{hyperref}
\usepackage{cleveref}

\newcommand{\SL}{\mathop{\rm SL}}

\newcommand{\SU}{\mathop{\rm SU}}

\def\p{\partial}

\def\bi{\begin{itemize}}
\def\ei{\end{itemize}}
\def\be{\begin{equation}}
\def\ee{\end{equation}}
\newcommand{\bea}{\begin{eqnarray}}
\newcommand{\eea}{\end{eqnarray}}

\title{Emergent Gravity from the
Entanglement Structure in Group Field
Theory}

\author[a]{Jinglong Liu,}
\author[b]{Stephon Alexander,}
\author[a,c,d]{Antonino Marcian\`o}
\author[e]{and Roman Pasechnik}
\affiliation[a]{Center for Field Theory and Particle Physics \& Department of Physics, Fudan University,
200433 Shanghai, China}
\affiliation[b]{Brown Theoretical Physics Center and Department of Physics,
Brown University,
RI 02912, USA}
\affiliation[c]{Laboratori Nazionali di Frascati INFN,
Frascati (Rome), Italy, EU}
\affiliation[d]{INFN sezione Roma Tor Vergata,
I-00133 Rome, Italy, EU}
\affiliation[e]{Department of Physics, Lund University, SE-223 62 Lund, Sweden, EU}

\emailAdd{jlliu21@m.fudan.edu.cn}
\emailAdd{stephon\_alexander@brown.edu}
\emailAdd{marciano@fudan.edu.cn}
\emailAdd{roman.pasechnik@fysik.lu.se}

\abstract{We couple to group field theory (GFT) a scalar field that encodes the entanglement between manifold sites. The scalar field provides a relational clock that enables the derivation of the Hamiltonian of the system from the GFT action. Inspecting the Hamiltonian, we show that a theory of emergent gravity arises, and that this can be recast according to the Ashtekar's formulation of general relativity. The evolution of the GFT observables is regulated by the Shr\"odinger equation generated by the Hamiltonian. This is achieved by imposing a renormalization group (RG) flow that corresponds to a simplified Ricci flow. As a consequence of the quantization procedure, the Hamiltonian is recovered to be non-Hermitian, and can be related to the complex action formalism, in which the initial conditions and the related future evolution of the systems are dictated by the imaginary part of the action.}

\begin{document}
\maketitle
\flushbottom

\section{Introduction}
\noindent
The Einstein's theory of general relativity (GR), which shows that gravity is generated by the dynamics of geometry \cite{https://doi.org/10.1002/andp.19163540702}, has been successfully and thoroughly scrutinized in many ways and totally embraced in modern physics for studies of macroscopic dynamics. At microscopic scales, strong and electroweak interactions become predominant, described by renormalized quantum field theories (QFTs), and reconciling them to a theory of gravity is still a mystery. The difficulties in quantizing gravity lie in the foundations of the two theories: GR relies on the dynamics of spacetime, without any preferred reference frame, while quantum theories of fundamental forces require a fixed background with a clear distinction between space and time.

A crucial question in quantizing gravity is not only how to quantize the spacetime, but mainly what we mean by that. Regge calculus \cite{Regge:1961px} provides a discretized formalism for GR, in which a triangulation of 4D spacetime is used and the curvature is described by the deficit angles associated with the faces shared by the glued 4-simplices. Hinging on its deep relation to the Regge calculus, a discretized model called spin foam has been created \cite{Baez:1997zt,Baez:1999sr}, providing a quantum history of spin networks. A spin foam is an quantum gravity analogue of Feynman diagrams used in other gauge theories, and consists of a set of rules to compute the amplitudes for the faces, edges and vertices in the spin foam. 

It was found in \cite{DePietri:1999bx} that the Barrett–Crane spin foam model can be interpreted as a Feynman graph of the Boulatov-Ooguri type field theory \cite{Boulatov:1992vp,Ooguri:1992eb}, the latter being nothing more than a topological BF theory. A new type of theory called group field theory (GFT) was introduced in the literature. A general GFT can be understood as the combination of tensor models \cite{DePietri:2000ii} for gravity and spin foam models.

Furthermore, the analog gravity approach \cite{barcelo2011analogue} is a theoretical scenario that concerns retrieving an emergent gravitational field from other physical systems, for example from electromagnetic systems and lattice models. Analog gravity provides a novel perspective on gravity as not being a fundamental force of Nature opening the pathway to shape new strategies and develop new applications. In particular, one can consider gravity itself as originating from another microscopic system at the fundamental level, e.g.~a QFT. This approach is at the origin of the paradigm of induced or emergent gravity \cite{Sakharov:1967pk,Visser:2002ew}, i.e.~considering the gravitational interactions as an emergent phenomenon in the infrared regime of the underlined QFT. Another perspective towards emergent gravity is called entropic gravity \cite{Carroll:2016lku}, in which there is a deep relation between gravity and entropy from quantum entanglement. A more recent scenario suggests that gravity arises from quantum information \cite{qi2018does}. Specifically, besides emergent gravity originating from quantum entanglement, this scenario also implies that gravity can emerge from quantum error-correction code (QECC). In \cite{bain_spacetime_2020}, for instance, the anti-de Sitter space and its boundary conformal field theory are both emergent from erasure-protection QECC. Hence, AdS/CFT correspondence itself becomes emergent, with local bulk degrees of freedom encoding the boundary degrees of freedom redundantly. The relation between emergent gravity and tensor networks can be also recovered in this framework. Nonetheless, tensor network is still a toy model, lacking several crucial ingredients of quantum gravity including an implementation of dynamics. Random tensor networks have already been related to Regge calculus in \cite{han2017discrete} in three dimension, and tensor network states have been connected to GFT entanglement graph states via projected entangled-pair states in arbitrary dimension \cite{Colafranceschi:2020ern}. We may then ask at this stage if and how gravity is emergent from the GFT entanglement structure.

A well known problem in quantum gravity is to consistently describe time and its origin \cite{Isham:1992ms,anderson2012problem}. This corresponds, in quantum gravity, to the failure of the de-parametrization process to recover time. Within the ADM formulation of GR \cite{PhysRev.116.1322} the scalar part of the Hamiltonian constraint instantiates gauge transformations for time-like diffeomorphisms, while the vector part of the Hamiltonian constraint generates space-like diffeomorphisms. Within the Hamiltonian non-perturbative approaches to quantum gravity including loop quantum gravity \cite{rovelli2004quantum}, physical states are annihilated by the Hamiltonian constraint; gauge-invariant observables must be invariant under time and space re-parametrizations. Relational observables \cite{Rovelli:1990ph,marolf1995almost,gambini2001relational,rovelli2002partial,dittrich2006partial} that have a relational evolution with respect to one another have been introduced for this purpose --- nonetheless, it is still debated if the relational evolution can actually solve the problem of time. Within GFT, a coupled scalar field can be used to provide either effective dynamics in cosmology \cite{Oriti:2016qtz,Oriti:2016ueo,Li:2017uao}, or a relational clock \cite{Wilson-Ewing:2018mrp,Gielen:2020fgi}. Noteworthy, the Hamiltonian and Friedman equations naturally originate in this emergent scheme.

Combining the aforementioned points, in this work we consider gravity to emerge from GFT with entanglement structure and a relational clock, hence evading the problem of time. We foresee the model to provide as well a link between the dynamics in gravity and in tensor networks. We will start with a brief introduction of related theoretical bases, including the GFT in Sec.~\eqref{SecGFT}, the emergent gravity model proposed by \cite{Lee:2019uix} in Sec.~\eqref{LeeEmergent} and the renormalization group (RG) flow -- in Sec.~\eqref{SecRG}. In Sec.~\eqref{SecHamil}, we will construct the action of the theory from the GFT action, and reproduce a Hamiltonian function similar to that studied in \cite{Lee:2019uix}, whose dimension, topology and geometry are all dynamical. In Sec.~\eqref{SecEvo}, we will calculate the inner product between two GFT entanglement graph states, as derived by the path-integral approach. On the other hand, we will show that the path-integral provides a particular RG flow whose RG equation is the Shr\"odinger equation. We will show how the Hamiltonian can be regarded as a generator of RG flow, and that an emergent mass scale arises in the Hamiltonian, making it possible to exploit quantum mechanics to introduce renormalization in quantum gravity, in a simplified version of the Ricci flow \cite{topping2006lectures}. We will then reproduce the Ashtekar formalism of GR \cite{PhysRevLett.57.2244}. We will finally comment, in the concluding remarks of Sec.~\eqref{Conclu}, that evolution of the GFT entanglement states provides a complex action, and that our model hence provides an example of the future-included complex theory \cite{Nagao:2017ztx,nagao2022reality}.

\section{Group Field Theory formalism}\label{SecGFT}

The GFT is a QFT on a base manifold that is a Lie group and whose structure has been shown -- see e.g.~Ref.~\cite{Krajewski:2011zzu} for a review -- to correspond to the discretization of space-time manifolds. Thus, GFT can be connected to background independent formulations of quantum gravity, including loop quantum gravity \cite{rovelli2004quantum}, the spin foam formalism \cite{perez2004introduction} and the causal dynamical triangulation \cite{Ambjorn:2012jv}. On a $d$-dimensional sub-manifold of a $D$-dimensional manifold, with $D=d+1$, we may introduce the triangulation as a tessellation procedure that cuts the space into cells, each one being homeomorphic to a $d$-dimensional simplex ($d$-simplex). Dual complexes to the $d$-simplices are 2-skeletons. Each $d$-simplex corresponds to a vertex on the dual complex, while each $(d-1)$-simplex that belongs to the $d$-simplex provides a semi-edge incident to the vertex of the dual complex. For each vertex dual to a $d$-simplex, there are $D=d+1$ $(d-1)$-simplices. This provides a pathway to represent the $d$-simplex in the $D$-dimensional manifold: each one of the $D$ $(d-1)$-simplices of the $d$-simplex is representing a dimension of the $D$-dimensional manifold. A field can be then introduced as defined on $D$-copies of the group manifold. Each group element then corresponds to one semi-edge, which in turn is dual to one of the $D$ $(d-1)$-dimensional faces of the $d$-simplex,
\begin{equation}
\begin{aligned}
    \Phi: \quad G^D  &\longrightarrow \mathbb{C}\,,\\
    \quad \mathbf{g} &\longmapsto \Phi(\mathbf{g})\,,
\end{aligned}
\end{equation}
where $\mathbf{g}=\{g_1, g_2, \cdots g_i, \cdots g_D\}$, each $g_i$ corresponding the ``color'' of a semi-edge $i$. The field, which enjoys the symmetry $\Phi(g_i)=\Phi(g_i g)$ for any $g$, describes a $d$-simplex in a discretized space --- the color $i$ labels the $D$ $(d-1)$-dimensional faces of the simplex, each one being assigned a group element $g_i$ --- as represented on the $D$-manifold. These are vertices of the interaction corresponding to fundamental space-time events. The dynamics of GFT can be described by an action \cite{oriti2006group}
\begin{equation}
    S[\Phi]=K[\Phi]-V[\Phi]\,,
\end{equation}
where $K[\Phi]$ is the kinetic term,
\begin{equation}
    K[\Phi]=\frac{1}{2}\Bigl(\prod_i^D\int dg_id\tilde{g}_i\Bigr)\Phi(g_i)\mathcal{K}(g_i\tilde{g}_i^{-1})\Phi(\tilde{g}_i) \,.
    \label{GFTKterm}
\end{equation}
The two copies of the field in Eq.~\eqref{GFTKterm} individuate a $d$-simplex, as it is shared by the two different $D$-simplices represented by each $\Phi$ (space-time events). The kernel $\mathcal{K}$ then describes the propagation of the geometric degrees of freedom, the $D$ $(d-1)$-simplices belonging to a $d$-simplex, from one vertex of interaction to another. Above, $V[\Phi]$ is the interaction vertex encoding how $d$-simplices, each one belonging to a distinct $D$-simplex, are glued together to form a collective $D$-simplex. The group elements associated to the $D$ $(d-1)$-simplices (combined in a $d$-simplex) can be labelled with indices $i',j'=1,\dots D+1$, with $i'\neq j'$. Removing prime in the indices for simplicity of notation, a possible example of interaction potential is provided by the expression:
\begin{equation}
    \frac{\lambda}{(D+1)!}\Bigl(\prod_{i\neq j=1}^{D+1}\int dg_{ij}\Bigr)\Phi(g_{1j})\cdots\Phi(g_{(D+1)j})\mathcal{V}(g_{ij}g_{ji}^{-1}) \,.
\end{equation}
The choice of $\mathcal{K}$ and $\mathcal{V}$ determines the theory. It is convenient to transform the field $\Phi(\vec{g})$ in the spin-intertwiner representation using the Peter-Weyl expansion:
\begin{equation}
\Phi(\vec{g})=\sum_{j_i,m_i,n_i,\iota}\phi^{\vec{j},\iota}_{\vec{m}} \mathcal{I}^{\vec{j},\iota}_{\vec{n}}\prod_{a=1}^{d+1}\frac{1}{d(j_a)}D_{m_an_a}^{j_a}(g_a) \,,\label{PeterWeyl}
\end{equation}
where $\mathcal{I}^{\vec{j},\iota}_{\vec{n}}$ denotes an intertwiner, which is associated to a vertex (dual to a $d$-simplex, i.e.~a tetrahedron in $D=4$) and labeled by $\iota$, $j$ is the spin of the irreducible representation (irrep) of the group element associated to different edges (dual to the ``faces'' of the triangulations, i.e. the $(d-1)$-simplex), $m$ and $n$ are the related magnetic numbers of the irrep, $\{j_i\}$ are a set of spins for $i=1,\dots D$, and similarly $\{m_i\}$ and $\{n_i\}$, and $D_{mn}^j(g)$ is a representation of the group element $g$, the Wigner matrix --- see e.g.~\cite{makinen2019introduction}.

Finally, the partition function of a GFT is defined by
\be 
Z=\int \mathcal{D}\Phi e^{-S[\Phi]}=\sum_\Gamma\frac{\lambda^N}{\rm sym(\Gamma)}Z(\Gamma)\,,
\ee 
where $\Gamma$ denotes a Feynman diagram, the links of which are composed by $D$ symmetrized strands (the $D$ group elements in the argument of the field), and the vertices of which have the combinatorial structure of a $D$-simplex, $N$ is the number of its vertices, and finally sym($\Gamma$) is a symmetry factor associated to the graph.

\subsection{GFT coupled with scalar fields}

Coupling the GFT with scalar fields has been studied in the literature \cite{Li:2017uao, Wilson-Ewing:2018mrp, Gielen:2020fgi}. In this context, the scalar fields act as gauge fields, and may play the roles of relational clock or space directions. We may focus on the case in which only one scalar field $\chi$ is coupled in GFT, and the translation and reflection symmetries are satisfied. The action is then invariant under the transformations $\chi\rightarrow \chi+\alpha$ and $\chi\rightarrow -\chi$, as is studied in \cite{Wilson-Ewing:2018mrp}. The scalar fields $\chi$ and $\chi'$ for the two connected simplices are integrated over in the action. The Peter-Weyl expansion of the action recasts 
\begin{equation}
    S[\phi]=\frac{1}{2}\sum_J\int d\chi \sum_{n=0}^{\infty} \phi_J(\chi)\mathcal{K}_J^{(2n)}\p_\chi^{(2n)}\phi_J(\chi)\,,
\end{equation}
where the collective index $J=(j, m ,\iota)$ has been introduced, $\chi'$ has been written as $\chi+\delta\chi$, and $\phi(\chi')$ has been Taylor expanded around $\chi$, hence involving infinite functional derivatives in $\chi$, and the integral over $\delta\chi$ has provided the definition of the kernel
\begin{equation}
   \mathcal{K}_J^{(2n)}=\int d\delta\chi\frac{1}{(2n)!}\mathcal{K}_J((\delta\chi)^2)((\delta\chi)^2)^n\,.
\end{equation}
The potential term is ignored here. Keeping terms with $n=0,1$ only, the conjugate momentum of $\Phi(\chi)$ is found to be
\begin{equation}
    \pi_J(\chi)=\frac{\delta S[\phi]}{\delta\bigl(\p_{\chi}\phi_J(\chi)\bigr)}=-\mathcal{K}_J^{(2)}\frac{\p\phi_J(\chi)}{\p\chi}\,,
\end{equation}
and the corresponding Hamiltonian is derived by the Legendre transform
\begin{equation}
    \mathcal{H}=-\sum_J\biggl[\frac{\pi_J(\chi)^2}{2K_J^{(2)}}+K^{(0)}_J\frac{\phi_J(\chi)^2}{2}\biggr]\label{MultiScalarH}\,.
\end{equation}
After second quantization, the Hamiltonian that can be decomposed into ladder operators and its eigenstates evolve according to the Shr\"odinger equation. The canonical formalism then generates an effective cosmology, providing the first Friedmann equation.

\subsection{GFT entanglement graph state}\label{GFTentangleState}

The quantum state of GFT is an element of the Fock space,
\begin{equation}
    \mathcal{F}=\bigoplus_{V=1}^\infty {\rm sym}(\mathcal{H}^1\bigotimes\cdots \bigotimes\mathcal{H}^V) \,,
    \label{FockSpace}
\end{equation}
constructed by the Hilbert spaces $\mathcal{H}^v$ for each vertex $v\in V$. In the presence of an interaction term in the action, correlations between the vertices should emerge, meaning that different vertices will be entangled. Entanglement for GFT quantum states, selected by the projection operator,
\begin{equation}
    \mathbb{P}_i^{x\bigotimes y}:=\int dh_i^{xy} dg_i^x dg_i^y |g_i^x\rangle\langle g_i^x h_i^{xy}|\bigotimes |g_i^y\rangle\langle g_i^y h_i^{xy}|\label{linkmap}\,,
\end{equation}
was delved in \cite{Colafranceschi:2020ern}. In Eq.~\eqref{linkmap}, $h_i^{xy}$ denotes the group element that renders the state $|g^i_x\rangle$ invariant by right action, called gluing element, with the labels denoting that it is integrated out to connect vertices $x$ and $y$ through the color $i$. For a state $|\psi\rangle \in \mathcal{H}^V$, the projector acts
\begin{equation}
    \mathbb{P}_i^{x\bigotimes y}|\psi\rangle =\int \prod_x d\mathbf{g}^x\int \Bigl(\prod_{l\in \gamma}dh_l\Bigr)\psi(\cdots, g_i^xh_l,\cdots, g_i^yh_l,\cdots)\bigotimes_x|\mathbf{g}^x\rangle \,.
    \label{LinkonState}
\end{equation}
Integrating the functional $\psi$ over $h_l$ provides a convolution and results into having a link between the two sites $x$ and $y$. Then, for a full graph state, one may recover 
\be 
|\psi_\gamma\rangle=\prod_{x<y}\prod_{A_{xy}^i=1}\mathbb{P}_i^{x\bigotimes y}|\psi\rangle=\int \prod_x d\mathbf{g}^x\int \prod_{x<t_i(x)}d\mathbf{h}^{x\mathbf{t}(x)}\psi({g_i^xh_i^{xt_i(x)}})\bigotimes_x|\mathbf{g}^x\rangle \,,
\ee
where $d\mathbf{h}^{x\mathbf{t}(x)}:=dh_1^{xt_1(x)}\cdots dh_d^{xt_d(x)}$, and $t_i(x)$ is the target vertex of the link labeled by the color $i$ at the vertex $x$. The second product in the first line is over a matrix $A_{xy}^i$ called adjacency matrix, which is such that $A_{xy}^i=1$ only if the vertices $x$ and $y$ are connected through the link with color $i$, otherwise it vanishes. The Hilbert space for $\psi_\gamma$, supported on a graph $\gamma$ with $V$ vertices, is a subset of $H^V$.

To remove labels, we may sum over all the permutations $\pi$, hence finding
\be 
\psi_\Gamma(\mathbf{g}^1,\cdots,\mathbf{g}^V)={\rm sym}_x(\psi_\gamma({g_i^x(g_i^y)^{-1}}))=\sum_\pi\int\prod_{x<t_i(x)}d\mathbf{h}^{x\mathbf{t}(x)}\psi_\gamma(\{g_i^{\pi(x)}h_i^{xt_i(x)}\})\,.
\ee 
This operation can also be denoted as 
\be 
|\psi_\Gamma\rangle =\sum_{\pi\in S_v}\mathbb{P}_\pi|\psi_\gamma\rangle=\mathbb{P}_{\rm inv_\pi}|\psi_\gamma\rangle\,,
\ee 
where $\mathbb{P}_\pi$ is the permutation operator with permutation $\pi$.

The second quantization procedure finally leads to
\begin{equation}
    |\psi_\Gamma\rangle=\int \prod_xd\mathbf{g}^x\psi_\gamma(\mathbf{g}^1,\cdots,\mathbf{g}^V)\prod_x\phi^\dagger(\mathbf{g}^x)|0\rangle\,.
\end{equation}
The functional $\psi_\gamma$ can also be written as a product of functionals at different vertices, which means $\psi_\gamma(\mathbf{g}^1,\cdots,\mathbf{g}^V)=\prod_xf_x(\mathbf{g}^x)$, with the connectivity being provided by the integral over the gluing elements on the vertex weighted functional. In order to distinguish among different vertices, the scalar fields can be coupled to the graph, peaking correspondingly the functionals $f_x$.

\section{The renormalization group flow}\label{SecRG}

While integrating out the degrees of freedom at large energy scales, and hence rescaling the momentum, the RG flow provides the evolution (in energy scale) of operators in a given physical theory. The first step of the RG is the coarse-graining: the spacing between lattice sites can be averaged over a block, hence a UV cut-off $\Lambda$ can be rescaled to a $b\Lambda$, with $b<1$. Accordingly, the momentum is rescaled to $k\rightarrow k/b$. This procedure was first proposed as a block-spin RG in \cite{Kadanoff:1966wm}. Its continuum limit entails a differential equation called the RG equation, which yields the Wilsonian RG \cite{Wilson:1973jj,Wilson:1983xri}. The path-integral approach of the Wilsonian RG was clarified in \cite{Polchinski:1983gv}. The couplings in the RG depend on an arbitrary energy scale $\mu$, called sliding or renormalization scale. The behavior of the couplings given by a first-order differential equation determined by the $\beta$-function is known as the RG flow.

An infinitesimal version of the Wilsonian RG equation for the effective action is provided by the Polchinski equation \cite{polchinski1984renormalization,gilkey2018invariance}, with the RG 'time' defined by $t={\rm ln}\Lambda$. The Polchinski equation can be recast in the formalism of the heat equation \cite{widder1976heat}, provided that a heat flow is considered, revealing a close relation between the RG equation and the heat equation. Indeed, it is well known that the heat equation can be written in the form:
\begin{equation}
    \frac{\p f}{\p t}=\Delta f \,,
\end{equation}
where $\Delta$ denotes the Laplacian operator. This equation is analogous to the Shr$\ddot{\rm o}$dinger equation by setting an imaginary time $\tau=it$, which is used also in the diffusion Monte-Carlo method to solve the Shr$\ddot{\rm o}$dinger equation. Based on these considerations, it is meaningful to use the Shr$\ddot{\rm o}$dinger equation defined on the relational clock as the RG equation, as we will see in greater detail in the next sections.

\subsection{``Quantum renormalisation group"}\label{SecQRG}

Holographic duality, according to which the link between the boundary and the bulk is instantiated by the RG flow, has been playing over the last decades a crucial role in quantum gravity studies. However, the RG expected by holography theories is not captured by the Wilsonian RG, since the single-trace operators also determine the flow of multi-trace operators \cite{Heemskerk:2010hk,Faulkner:2010jy}. In Refs.~\cite{Lee:2011wx,Lee:2012xba,Lee:2013dln}, the Wilsonian RG was promoted to the quantum RG, in which single-trace operators will generate multi-trace ones. While in \cite{Lee:2016uqc}, it was proved that a RG is equivalent to a series of wave function collapses, which describe the evolution from a state associated with an action in field theory to an IR fixed point. We may delve into the main features of the quantum RG following Ref.~\cite{Lee:2016uqc}.

The wave function evolving in the quantum RG is associated to an action of a field theory, and it is defined in the space of single-trace operators. We will call such wave functions $S$-function, or $S$-state for the corresponding quantum state.

According to the partition function of a field theory, one can define a state as 
\begin{equation}
    |S\rangle = \int \mathcal{D}\phi e^{-S[\phi]}|\phi\rangle \,,
\end{equation}
with $|\phi\rangle$ denoting the basis' elements that span a Hilbert space, satisfying
\begin{equation}
    \langle \phi'|\phi\rangle = \prod_{ia}\delta(\phi'_{ia}-\phi_{ia}) \,,
\end{equation}
where $i$ labels a spacetime index or a vertex in a lattice model, and $a$ is the internal color of the degrees of freedom. In general, an action can be represented as 
\begin{equation}
    S=-\sum_n(J^{i_1\cdots i_n}O_{i_1\cdots i_n}) \,,
\end{equation}
where $O_{i_1\cdots i_n}$ is an operator depending on $n$ vertices, and $J$ is the corresponding source. If there is a symmetry, i.e.~there is a unitary operator $\hat{U}$ transforming the states as $\hat{U}|\phi\rangle=|U\phi\rangle$, such that $S[U\phi]=S[\phi]$, the action can be rewritten as a sum over singlet operators, which means
\begin{equation}
    S = -J^M O_M \,.
\end{equation}
There is a minimal set of operators $\{O_n\}$ that generates all the singlet operators, provided that \cite{Lee:2013dln}
\begin{equation}
    O_M = \sum_k c^{n_1\cdots n_k}_M O_{n_1}\cdots O_{n_k} \,,
\end{equation}
Here, the operators $O_n$ that compose the minimal set are the single-trace operators. As a result, the $S$-state can be recast as 
\begin{align}
    |S\rangle &= \int \mathcal{D}\phi e^{-J^MO_M}|\phi\rangle\notag\\
    &=\int \mathcal{D}\phi e^{\sum_k J^{n_1\cdots n_k}O_{n_1}\cdots O_{n_k}}|\phi\rangle \notag\\
    &= \int \mathcal{D}j\mathcal{D}\phi e^{-j^n(j_n^*-O_n)+\sum_k J^{n_1\cdots n_k}j^*_{n_1}\cdots j^*_{n_k}}|\phi\rangle \,,
\end{align}
where in the last line we have used the identity
\begin{equation}
    f(O)=\frac{1}{\pi}\int dj dj^* e^{-j(j^*-O)}f(j^*)\,,\label{idenJ}
\end{equation}
which has been proved in \cite{Lee:2011wx}, and the constant prefactor has been omitted. In \eqref{idenJ}, $j$ and its complex conjugate $j^*$ are auxiliary fields, labeled by $n$ and with measure $\mathcal{D}j=\prod_n dj_n dj_n^*$. The $S$-state is then represented by 
\begin{equation}
    |S\rangle = \int \mathcal{D}j \Psi[j]|j\rangle \,,
\end{equation}
where 
\begin{equation}
    |j\rangle = \int \mathcal{D}\phi e^{j^nO_n}|\phi\rangle\label{SingletraceS}
\end{equation}
is the basis constructed by single-trace operators, and 
\begin{equation}
    \Psi[j]=e^{-j^nj_n^*+\sum_k J^{n_1\cdots n_k}j^*_{n_1}\cdots j^*_{n_k}} \,.
\end{equation}
These results show that using the $S$-state, one can define the quantum states in the space of single-trace operators, and that the Wilsonian RG with all singlet sources can be promoted to a RG defined in the space involving only single-trace sources, named quantum RG in Refs.~\cite{Lee:2011wx,Lee:2012xba,Lee:2013dln,Lee:2016uqc}. 

We now inspect how the inner product between two $S$-states forms a coarse-graining in the RG, flowing a state from UV to IR.

Consider two $S$-states, 
\begin{align}
    |S_1\rangle = \int d\phi e^{-S_1[\phi]}|\phi\rangle \,,\\
    |S_2\rangle = \int d\phi e^{-S_2[\phi]}|\phi\rangle \,,
\end{align}
the inner product among which is provided by 
\begin{equation}
    Z = \langle S_1|S_2\rangle = \int \mathcal{D}j'\mathcal{D}j\Psi[j']^*\Psi[j]\langle j'|j\rangle \,,
\end{equation}
with $|j\rangle$ denoting the single-trace state defined in Eq.~\eqref{SingletraceS}. We only need to consider the overlap between single-trace states. To evaluate the inner product $Z$, an example for real scalar field $\phi$ is provided by Ref.~\cite{Lee:2016uqc}, in which 
\begin{equation}
    Z = \int d\phi e^{-S_1-S_2}\,,
\end{equation}
with $S_1=\frac{m^2}{2}\phi^2$ denoting a reference action with mass $m$. To apply the coarse-graining, an auxiliary field $\Phi$ with mass $\mu$ is introduced, which means the reference action is shifted to $S=\frac{m^2}{2}\phi^2+\frac{\mu^2}{2}\Phi^2$. By redefining the fields $\phi=\phi'+\Phi'$ and $\Phi=\frac{m}{\mu\sqrt{2dz}}(-2dz\phi'+\Phi')$, with $\Phi'$ a small fluctuation, the action $S_1+S_2$ becomes
\begin{equation}
    S=\frac{m^2e^{2dz}}{2}\phi'^2+\frac{m^2}{4dz}\Phi'^2+S_2[\phi'+\Phi']\,.
\end{equation}
The action $S_2$ can be then expanded around $\phi'$. We keep the lowest two orders, and then sum over $\Phi'$ performing a Gaussian integral. Hence, we rescale the field $\phi'\rightarrow e^{-dz}\phi''$ and redefine $\phi''\rightarrow \phi$. Two corrections are finally introduced, respectively,
\begin{align}
    &\delta_1S_2 = -\frac{dz}{m^2}\biggl(\frac{\partial S_2[\phi']}{\partial \phi'}\biggr)^2 \,,\label{d1S1}\\
    &\delta_2S_2 = -dz \phi \frac{\partial S_2[\phi]}{\partial \phi}\label{d2S1}\,.
\end{align}
The action finally becomes 
\begin{equation}
    S = \frac{m^2}{2}\phi^2 + S_1[\phi] + \delta_1S_1[\phi]+\delta_2S_2[\phi] \,.\label{AcexQRG}
\end{equation}
In Eqs.~\eqref{d1S1}-\eqref{d2S1}, $S_2[\phi']$ and $S_2[\phi]$ appear in a similar way, respectively, in $\phi'$ and $\phi$. The conjugated momentum to $\phi$ can be then recovered by considering the variation
$$\pi = \frac{\delta S[\phi]}{\delta (d\phi/dz)} = dz \frac{\delta S[\phi]}{\delta \phi}\,.$$ 
As a result, the inner product becomes
\begin{equation}
    Z = \langle S_1 | e^{-dz\hat {H}}|S_2\rangle \,,\label{PartiexamQRG}
\end{equation}
where the Hamiltonian operator acquires the form
\begin{equation}
    \hat{H}=\frac{1}{m^2}\hat{\pi}^2+i\hat{\phi}\, \hat{\pi}\,,
\end{equation}
with $\hat{\phi}$ and $\hat{\pi}$ denoting the operators corresponding to the field $\phi$ and $\pi$ and satisfying the symplectic relation $[\hat{\phi},\hat{\pi}]=i$. \\

This result implies that the inner product between any $S$-state $|S_2\rangle$ and the reference $S$-state $|S_1\rangle$ can be regarded as a coarse-graining from another state with action $S_1'=S_1+\delta S$. The deformation $\delta S=-dz H$ results from integrating out a small fluctuation of the scalar field, and is generated by the Hamiltonian. In the shifting and rescaling of the field $\phi$ and in the deformation of the action, one can observe that the reference action $S_1$ does not change. Therefore, $|S_1\rangle$ individuates the ground-state of the operator $\hat{H}$. Unlimited copies of the operator $e^{-dz \hat{H}}$ can be then inserted in the inner product $\langle S_1|S_2\rangle$, in order to further deform the action.  

For a series of independent Hamiltonians $h_i$ corresponding to different vertices, one can consider a Hamiltonian $H=\sum_i\alpha_i h_i$, where $\alpha_i$'s are constants that determine the speed of coarse-graining at each vertex --- for simplicity, we choose here $\alpha_i=1$. The reference state $|S_0\rangle$ is a direct product of the ground-states of $h_i$'s. Due to the insertion of the coarse-graining operator $e^{-dz\hat{H}}$, the partition function becomes 
\begin{equation}
    Z= \int \mathcal{D}j^{(0)} \Psi[j^{(0)}]\langle S_0|e^{-dz\hat{H}}|j^{(0)}\rangle\,,\label{QRGpartition}
\end{equation}
where the evolution of $|j^{(0)}\rangle$ is governed by
\begin{equation}
    e^{-dz\hat{H}}|j^{(0)}\rangle = \int \mathcal{D}\phi\, e^{j^{(0),n}O_n-dz\sum_k c^{n_1\cdots n_k}[j^{(0)}]O_{n_1}\cdots O_{n_k}}|\phi\rangle\,.\label{QRGevolve}
\end{equation}
The Hamiltonian introduces multi-trace operators by deforming the original action constructed in terms of polynomials of single-trace operators. As a result, for a system with multiple vertices, the ground-state of which is determined by local Hamiltonians, the $S$-states that involve only single-trace operators generate multi-trace operators in the evolution. The Hamiltonian removes the entanglement present in the UV state while gradually approaching the IR fixed point provided by the reference state.

\section{Lee's emergent gravity theory}\label{LeeEmergent}

In Ref.~\cite{Lee:2019uix}, S.S. Lee proposed a theory of emergent gravity based on a lattice model, the Hamiltonian of which is the generator of the quantum RG introduced in last section. We briefly review the simplified version of that theory in relation to our model.

In Lee's model, the fundamental quantities are defined on a set of $L$ sites, to each of which is assigned the same number of local degrees of freedom. Hence, the variables are provided by a matrix $\Phi^A_x$, where $A$ labels the species called colors, and $x$ labels the sites. $\Phi^A_x$ is the eigenvalue of the operator $\hat{\Phi}^A_x$, whose eigenstate $|\Phi^A_x\rangle$ satisfies
\begin{equation}
    \langle \Phi^A_x|\Phi^B_y\rangle=\delta^{A,B}\delta_{x,y}\,. 
\end{equation}
The Hilbert space $\mathbb{H}$ is spanned by $\bigotimes_{x,A}|\Phi^A_x\rangle$. The conjugate momentum of $\Phi^A_x$ is denoted as $\Pi^x_A$, and satisfies the symplectic relation 
\begin{equation}
[\Phi^A_x,\Pi^y_B]=i\delta^A_B\delta^y_x\,.
\end{equation}
The Hilbert space $\mathbb{H}$ can be decomposed into Hilbert spaces at each vertex
\begin{equation}
    \mathbb{H}=\bigotimes_x\mathbb{H}_x \,,
\end{equation}
which is called a frame. The symmetry group of the colors is the orthogonal group $O(M)$, $M$ denoting the number of the colors. The elements of the basis of the Hilbert space can be then transformed to another frame according to $|\Phi^A_i\rangle \rightarrow |g^I_i \Phi^A_I\rangle \equiv |\tilde{\Phi}\rangle$, where $g^I_i\in \SL(L,\mathbb{R})$, with $L$ denoting the number of sites. The inner product among states that belong to different bases connected by a transformation, can be written as
\begin{equation}
    \langle \tilde{\Phi}|\tilde{\Phi}'\rangle = \prod_{i} \prod_{A}\delta(g^I_i(\Phi^A_I-\Phi'^A_I))=\prod_{I,A}(\Phi^A_I-\Phi'^A_I)\,,\label{Vertextrans}
\end{equation}
where $g^I_i=\delta^I_i$ has been used. This represents the basis transformation in the infinite dimensional Fock space.

The connectivity between sites are provided by the entanglement. On the other hand, the entanglement is encoded in collective variables \cite{Lee:2016uqc}, which can be used to calculate the quantum information between the entangled sites. Considering the connectivity between sites, the Hamiltonian constraints constructed out of $\Phi^A_i$ and $\Pi^i_A$ include two parts. The first part is the spacial diffeomorphism which transforms one vertex with label $i$ to another one with label $j$. Eq.~\eqref{Vertextrans} implies that the spacial diffeomorphisms can be realized resorting to the gauge group $\SL(L,\mathbb{R}$), the generators of which are the ones for ${\rm GL}(L,\mathbb{R})$, excluding dilatation. The generators for ${\rm GL}(L,\mathbb{R})$ are provided by
\begin{equation}
    \hat{\mathcal{G}}^x_{\ y}=\frac{1}{2}\bigl({\hat{\Pi}^x}_A{\hat{\Phi}^A}_y+{\hat{\Phi}^A}_y{\hat{\Pi}^x}_A\bigr) \,,
\end{equation}
while the $\SL(L,\mathbb{R})$ generators can be expressed as it follows 
\begin{equation}
    \hat{G}^x_{\ y} = \hat{\mathcal{G}}^x_{\ y} - \frac{1}{L}\hat{\mathcal{G}}^z_{\ z} \delta^x_{\ y}\,,
\end{equation}
where $\hat{\mathcal{G}}^z_{\ z}$ is the generator of dilatation. To encode the shift of each vertex, a new tensor $N^x_{s,y}$ called shift tensor need to be defined, where the label $s$ stands for `shift' and $x,y$ label a generic vertex. The contraction of the $\SL(L,\mathbb{R})$ generators with the shift tensor induces the shift of vertices to one another, hence realizing spatial diffeomorphisms. Taking the trace over the sites $x$ and $y$, the generator is expressed by
\begin{equation} \label{SDC}
    \hat{G}_s={\rm tr}\{\hat{G}N_s\}\,.
\end{equation}
The time-diffeomorphism constraint is instead expressed by
\begin{equation} \label{TDC}
    \hat{H}_l ={\rm tr}\{\hat{H}N_l\}
    ={\rm tr}\Biggl[\Biggl(-\hat{\Pi}\hat{\Pi}^T+\frac{\tilde{\alpha}}{M^2}\hat{\Pi}\hat{\Pi}^T\hat{\Phi}^T\hat{\Phi}\hat{\Pi}\hat{\Pi}^T\Biggl)N_l\Biggl] \,,
\end{equation}
where $\tilde{\alpha}$ is an arbitrary positive constant, and $N_l$ denotes lapse tensor. The diagonal part of $N_l$ provides a multi-fingered time evolution for the sites, while the off-diagonal parts enjoy frame rotational symmetry, which entails a superposition of 'clocks', and hence generates another set of them. In Eq.~\eqref{TDC}, the first term corresponds to a kinetic term, while the second one -- to a hopping term, which becomes $|\nabla \Phi|^2$ in the continuum limit.

When a reference state exists, i.e.~there is a state $|0\rangle$ invariant under the action of the Hamiltonian constraints provided above, the inner product between a generic state $|\chi\rangle$ and the reference state $|0\rangle$ may encode the insertion of an arbitrary number of copies of $e^{-i\epsilon(\hat{H}_l+\hat{G}_s)}$. An example of reference state is provided by the state $|0\rangle = \int \mathcal{D}\Phi |\Phi\rangle$, from the scalar product of which with $|\chi\rangle$ it is possible to recover the path-integral formula
\begin{equation}
    \langle 0|\chi\rangle=\int \mathcal{D}\Phi\mathcal{D}k \mathcal{D}v\mathcal{D}y e^{iS}\chi(\Phi,k) \,.\label{LeePI}
\end{equation}
In Eq.~\eqref{LeePI}, quantum states are written in the basis of the collective variables $k$, in turn provided by the auxiliary fields $j$ introduced in Eq.~\eqref{idenJ}, while $\chi(\Phi,k)$ denotes a generic function. The action reads
\begin{equation}
    S=-\int d\tau {\rm tr}[(-\phi(\tau)\p k(\tau)-N_l(\tau)H(\tau)-N_s(\tau)G(\tau)] \,,
    \label{LeeAction}
\end{equation}
with $\tau$ being defined in terms of $\epsilon$, by $d\tau=\epsilon$. The path-integral has the same expression as in Eq.~\eqref{QRGpartition}. The action in Eq.~\eqref{LeeAction} is derived by decomposing the state $|\phi\rangle$ in Eq.~\eqref{QRGevolve} into a basis of collective variables, hence generating a quantum RG, as introduced in Sec.~\eqref{SecQRG}. This is a simplified version of the Lee's model in \cite{Lee:2019uix}, without symmetry breaking of the colors into physical and redundant degrees of freedom. It carries nonetheless enough complexity (at the level of the action), which is enough for our analysis here. The spacetime metric is emergent from the collective variables $k$, and the saddle point equation of motion of the action \eqref{LeeAction} shows a solution of de-Sitter like space-time with Lorentzian signature. Within this theoretical framework, the manifold dimensions, the topology and the geometry are all emergent, being determined by the collective variables.

\section{Emergent Hamiltonian of gravity from GFT}\label{SecHamil}

A commonly accepted idea in the recent literature \cite{faulkner2013quantum, hubeny2007covariant,van2010building,maldacena2013cool,maldacena2016remarks,maldacena2016conformal,saad2018semiclassical,mertens2017solving,kitaev2019statistical,fukuma2003holographic,dieks2015emergence,banks2020holographic} is that spacetime and gravity are emergent from the entanglement structure of underlying microscopic quantities. Since GFT can describe the dynamics in a discretized space, and since entanglement plays a crucial role in relating different simplices, we can consider building emergent gravity models within the GFT regime. Specifically, in this section we build a quantum gravity model that can reproduce the essential Hamiltonian introduced by Lee in \cite{Lee:2019uix} and described in Sec.~\eqref{LeeEmergent}. Hence, most part of the results derived by Lee will still hold, while we demonstrate how such a theory of gravity emerges from GFT.

\subsection{Constructing the action}

We resort to the idea of entanglement-graph introduced in GFT at the vertices of which the action of the projection operator (\ref{linkmap}) is defined. Recall that within the GFT action, the potential term encodes how $d$-simplices connect to form a $(d+1)$-simplex. We can write the potential term as the product of all the projection operators acting on the vertices of a graph. Although the projector operator does not admit a functional representation, we can recover GFT entanglement-graph states in an alternative way. While evaluating the inner product between states, for instance in the path-integral formulation of the theory, we may indeed keep the first order of the interaction term and act with it on the GFT states, forming a GFT entanglement-graph state. The action of the kinetic term will then provide the evolution of these states. The information encoded in the action of the projection operator is included in the graph: it is described as an adjacency matrix $A_{xy}=1$ if $x$ and $y$ are connected, otherwise vanishing.

To make the entanglement physical, we assign ultra-local degrees of freedom (DOF) $\chi^i_x$ to each vertex, labelled by $x$, following a vertex permutation symmetry. These DOF then form a frame, whose internal color is denoted with $i$. The projection operator $\mathbb{P}_i^{x\bigotimes y}$ introduces entanglement between the vertices $x$ and $y$. We then expect a propagator between the DOF of the entangled vertices to be induced. Nonetheless, since the DOF assigned to different vertices must be considered as distinct ones, the propagators among them will vanish. Consequently, the couplings entering the expressions will no longer have a well defined meaning.

A better choice of the couplings arises from conceiving the variables coupled at each vertex to include non-local information. This latter can be provided by the correlations between the DOF assigned to every pair of vertices. The coupling to a vertex $x$ is then a massless local scalar field $\lambda_x=\chi_x^i\sum_y\chi_y^i$ --- $i$ labeling the internal color and all the repeated color indices being summed over. Consequently, any GFT quantum state or any functional that are defined on that vertex involve the correlations between all the pairs of vertices that include $x$. The quantum states or the functionals are still local, since the correlations are averaged over all the pairs, without any connectivity between vertices being introduced. If the connectivity is introduced by the projection operators, only the connected pairs described by $\lambda_{xy}=\chi^i_x\chi^i_y$ become the variables on the links of the graph, $\lambda_x$ now playing the role of a relational clock. In this case, the states of the Hilbert space $\mathcal{H}_x$ at a vertex $x$ can be defined for group elements that are extended to the field of the real numbers $R_x$. This latter is an inner product space, if we consider the inner product $\langle \cdot, \cdot\rangle: R_{x,1}^n\times R_{x,2}^n\rightarrow R_x$, where $R_{x,1}$ is a copy of the field of the real number associated to the variable $\chi^i_x$, $R_{x,2}$ is a copy of the field of the real number associated to the sum over all $\chi^i_{y\neq x}$, and we have written the extended Hilbert space defined on $x$ as $\mathbb{H}_x=\mathcal{H}_x\bigotimes R_x$.  A GFT quantum state is therefore an element of the extended Fock space
\begin{equation}
    \mathbb{F}=\bigoplus_{V=1}^\infty {\rm sym}(\mathbb{H}_1\bigotimes\cdots \bigotimes\mathbb{H}_V) \,.
\end{equation}
A generic state is then expressed by
\begin{equation}
    |\psi\rangle ={\rm sym}\biggl(\int \prod_x d\mathbf{g}_xd\mathbf{\lambda}_x\psi(\mathbf{g}_1,\mathbf{g}_2, \cdots,\mathbf{g}_V;\lambda_1,\cdots,\lambda_V)\bigotimes_x (|\mathbf{g}_x\rangle\bigotimes|\lambda_x\rangle)\biggr)\,, \label{GFTstate}
\end{equation}
where ``${\rm sym}$'' denotes symmetrization over all the permutations of the vertices. 

To introduce the propagator between local scalar fields, we further need to extend the projection operator into
\begin{equation}
        \mathbb{P}_i^{x\bigotimes y}:=\int d\lambda_x d\lambda_y dh_i^{xy} dg_i^x dg_i^y |g_i^x \rangle|\lambda_y\rangle \langle g_i^x h_i^{xy}|\langle \lambda_y|    \bigotimes |g_i^y\rangle|\lambda_x\rangle\langle g_i^y h_i^{xy}|\langle\lambda_x| \,.
        \label{Pop}
\end{equation}
Acting with this projection operator on a state \eqref{GFTstate}, a new term provided by $\langle \lambda_x|\lambda_y\rangle \langle \lambda_y|\lambda_x\rangle=|\langle \lambda_y|\lambda_x\rangle|^2$ is introduced as a multiplier in the functional. A term encoding the connectivity between two vertices then emerges. We will investigate the meaning of such terms in the path-integral part \eqref{SecEvo}.

Some remarks are necessary to clarify how the couplings are assigned. The ultra-local DOF $\chi^i_x$ defined on $x$ correspond to a local measurement of a field, which can be identified with a vector gauge field composed by a set of scalar fields. The coupling $\lambda_x$ includes the information of the correlations between $x$ and all the other vertices. It thus enables to introduce propagators of quantum fields, once projection operators have been defined. The connectivity between the vertices realizes causality among vertices, deciding if the propagator vanishes. While we expect the aforementioned objects, corresponding to the local values, the correlations and causality relations, to form a well-defined quantum field, it is not our purpose to analyse these structure here. We will rather focus on a model of gravity emerging from the local scalar field.

Using the relational clock, we then describe the dynamics of a single vertex. We note anyway that there is no physical meaning for the vertices $x$, since these are already symmetrized. We will provide the dynamics of an entire GFT entanglement graph (EG) state via path-integral in Sec.~\eqref{SecEvo}. As discussed in \cite{Wilson-Ewing:2018mrp}, coupled scalar fields may play a role of relational clocks. We can then resort to the same strategy, and regard our coupling value $\lambda_x$ as a relational clock that also encodes entanglement. The functional on the vertex $x$ becomes $\Phi(\mathbf{g}_x,\lambda_x)$. Following \cite{Wilson-Ewing:2018mrp}, the coupling of the scalar field in a GFT action should satisfy translation and reflection symmetry. The kernel $K$ in the GFT kinetic term \eqref{GFTKterm} must be then invariant under the transformation $\lambda\rightarrow \lambda+\epsilon$ and $\lambda\rightarrow -\lambda$. Nonetheless, since our couplings are constructed out of more fundamental (frame-)DOFs $\chi^i$, and $\chi_x^i$ and $\chi^i_y$ correspond to the same direction in the frame, it happens that the symmetries $\chi^i_x\rightarrow -\chi^i_x$ and $\chi^i_y \rightarrow -\chi^i_y$ hold simultaneously and hence leave the sign of $\lambda$ unchanged. As a result, the only symmetry that should be satisfied by $\lambda$ is the translational symmetry, and the kernel should be recast as $K(\delta \lambda)$.

In light of previous considerations, the action we consider then reads, for a single vertex, 
\begin{equation}
    S_x=\int d\mathbf{g}_x d\mathbf{g}'_x d\lambda_x d\lambda_x'\Phi(\mathbf{g}_x,\lambda_x)K(\mathbf{g}_x,\mathbf{g}'_x;\delta\lambda_x)\Phi(\mathbf{g}'_x,\lambda'_x)\label{initialaction}\,,
\end{equation}
where $\delta\lambda=\lambda'_x-\lambda_x$. As was mentioned above, we regard the coupling value $\lambda$ as a relational clock \cite{Wilson-Ewing:2018mrp, Gielen:2020fgi}, with respect to which the Hamiltonian is recovered. Since this coupling value includes non-local values, constructed by the product of the frame fields at different sites, it can be used to encode the information about the entanglement between different sites.

As we will see later in detail, the conjugate momenta of $\lambda$ enable to express the collective variables introduced in \cite{Lee:2019uix,lee2019state,Lee:2016uqc}, providing the mutual information between sites. This amounts to having the total entanglement between the site $x$ and other sites playing the role of 'clock' at the site $x$. The value of $\lambda_x$ does not provide the real time coordinate in physics, but its change does provide a flow in time, with different sites evolving at different rates.

The GFT functional can be finally expanded in the intertwiner space by using the Peter-Weyl transform \eqref{PeterWeyl}, with the intertwiner $\mathcal{I}$ and the Wigner matrices $D$ being integrated out by summing over $\mathbf{g}_x$ and $\mathbf{g}_x'$, which provides the identity. Then the action \eqref{initialaction} becomes 
\begin{equation}
    S_x=\sum_J\int d\lambda_x d\delta\lambda_x \varphi_J(\lambda_x)K(\delta\lambda_x)\varphi_J(\lambda_x+\delta\lambda_x)
    \label{NewAction}\,,
\end{equation}
where $\varphi_J$ is a functional mode within the action, and $J=\{\mathbf{j},\mathbf{m},\boldsymbol{}{\iota}\}$ is the collection of the representations $j$, the magnetic-numbers $m$ on the semi-links, the intertwiners $\iota$ as well as the vertices. We will later on omit the label $J$, following an assumption that only one mode in the Peter-Weyl transform is relevant for the effective dynamics \cite{Gielen:2020fgi}. Indeed, single modes dominate asymptotically in a wide range of models \cite{gielen2016emergence} besides being convenient for computational purposes.

\subsection{The Hamiltonian}

Using the action \eqref{NewAction}, a Hamiltonian can be finally derived from GFT. Expanding $\varphi_J(\lambda_x+\delta\lambda_x)$ around $\lambda_x$, we find 
\begin{equation}
    \varphi(\lambda')=\varphi(\lambda)+\sum_m\frac{1}{m!}\frac{\p^m\varphi}{\p\lambda^m}(\delta\lambda)^m\,.
\end{equation}
Keeping only the lowest three orders, the action becomes
\begin{equation}
    S_x=\int d\lambda_x\bigl( K^{(0)}\varphi_x^2+K^{(1)}\varphi_x\p_\lambda\varphi_x-\frac{1}{2}K^{(2)}\p_\lambda\varphi_x\p_\lambda\varphi_x\bigr)\label{finalaction}\,,
\end{equation}
where we have integrated by parts the last term, with the field $\varphi(\lambda_x)$ denoted as $\varphi_x$ for convenience, and defined 
\begin{equation}
    K^{(n)}=\int d(\delta\lambda) K(\delta\lambda)(\delta\lambda)^n\,.
\end{equation}
The conjugate momentum of $\varphi_x$ is 
\begin{equation}
    \pi_x \equiv \frac{\delta\mathscr{L}}{\delta(\p_\lambda\varphi)}=K^{(1)}\varphi_x-K^{(2)}\p_\lambda\varphi_x \,.
    \label{Notmomentum}
\end{equation}
Nonetheless, $\pi$ and $\varphi$ are not the real dynamical fields of our system. They are instead some collective properties depending on the background frame fields $\chi^i_x$.

Since the scalar field $\lambda_x$ is defined in an inner product space, the relation $\varphi_x=\sum_y\phi(\chi_x^i)\phi(\chi_y^i)$ is always true, and hence the dynamical fields in our system are $\phi_x^i=\phi(\chi^i_x)$, the conjugated momenta of which are, from Eq.~\eqref{finalaction},
\begin{equation}
    \omega^i_x=\pi_x\sum_{y\neq x}\phi^i_y\label{momentum}\,.
\end{equation}
Applying the Legendre transform to the Lagrangian, then solving for $\p_\lambda\varphi_x$ from Eq.~\eqref{Notmomentum}, $\p_\lambda\varphi=(K^{(1)}\varphi-\pi)/K^{(2)}$, and finally substituting into the Hamiltonian, we recover
\begin{equation}
    H_x=\biggl(-K^{(0)}-\frac{(K^{({1})})^2}{2K^{(2)}}\biggr)\varphi_x^2+\frac{\pi_x^2}{2K^{(2)}}+\varphi_x\pi_x\label{Hamiltonian1}\,.
\end{equation}
The terms in the parentheses individuate a mass term, the scale of which depends on the scale of the kernel $K(\delta\lambda)$. The second and the third terms then provide the dynamics of the system for a single site $x$. Using Eq.~\eqref{Hamiltonian1}, it is possible to reproduce the Hamiltonian constructed by Lee \cite{Lee:2019uix}. To prove this, we extract the quantity $\alpha \p_\lambda\varphi_x\p_\lambda\varphi_x$ out of the other terms. As a result, the Hamiltonian \eqref{Hamiltonian1} can be recast as 
\begin{equation}
    H_x=\biggl(-K^{(0)}-\Bigl(\frac{1}{2}+\alpha\Bigr)\frac{(K^{(1)})^2}{K^{(2)}}\biggr)\varphi_x^2-\bigl(\frac{1}{2}+\alpha\bigr)\frac{\pi_x^2}{K^{(2)}}
    +(1+2\alpha)\frac{K^{(1)}}{K^{(2)}}\varphi\pi+\alpha K^{(2)}\p_\lambda\varphi_x\p_\lambda\varphi_x\,.
\end{equation}
We interpret the second term as the kinetic term and the last term as the hopping term, and choose $\alpha=-1/4$ for which the two terms have opposite signs. We then exploit a Fourier transform for $\varphi_x$ and $\pi_x$,
\begin{equation}
    \begin{aligned}
        &\pi(\lambda_x)=\int \frac{dk}{2\pi} \pi(k_x)e^{-ik_x\lambda_x} \,, \\
        &\varphi(\lambda_x)=\int \frac{dk}{2\pi}\varphi(k_x)e^{-ik_x\lambda_x} \,,
    \end{aligned}
\end{equation}
and find the Hamiltonian in the momentum space
\begin{eqnarray}
    &&H(k_x)=-\frac{1}{4K^{(2)}}\pi_x(k_x)^2+\frac{K^{(2)}}{4}k_x\varphi(k_x)k_x\varphi(k_x)\label{NewHamiltonian} \\
   && +\frac{K^{(1)}}{2K^{(2)}}\varphi(k_x)\pi(k_x)+\frac{K^{(2)}}{4}\biggl(-\frac{4K^{(0)}}{K^{(2)}}-\frac{(K^{(1)})^2}{(K^{(2)})^2}\biggr)\varphi^2_x\,. \nonumber 
\end{eqnarray}
As we noticed previously, the scalar fields on different points correspond to different fields. Therefore, we define $k_x$ as the product of the conjugate momenta of $\chi^i_x$ and $\chi^i_y$, summing over $y$. Consequently, $k_x\propto \sum_y\omega_x^i\omega^i_y$, and we can define $k_x=l\sum_y\omega_x^i\omega^i_y$. The second term in \eqref{NewHamiltonian} finally becomes
\begin{equation}
    \frac{K^{(2)}l^2}{4}\sum_{y,z}\omega^i_x\omega_y^i\phi^j_y\phi^j_x\phi^k_x\phi^k_z\omega^t_z\omega^t_x \,. 
    \label{o4}
\end{equation}
The dynamics is provided by the propagator between two time-slices. The effective terms in Eq.~\eqref{o4} are those with $j=k$, and as a result the hopping terms in \cite{Lee:2019uix} are reproduced, namely,
\begin{equation}
    \frac{K^{(2)}l^2}{4}\sum_{y,z}\omega^i_x\omega_y^i\phi^j_y\phi^j_z\omega_z^k\omega_x^k \,.
    \label{Hoppingterm}
\end{equation}
From Eq.~\eqref{momentum} we recover two relations
\begin{align}
    &\pi_x^2=\frac{1}{\sum_y\phi_y^k\sum_z\phi_z^k}\omega^i_x\omega^i_x\,,
    \label{relation1}  \\
    &\varphi_x\pi_x=\phi^i_x\sum_y\phi^i_y\pi_x=\phi^i_x\omega^i_x \,.
    \label{relation2} 
\end{align}
Using \eqref{relation1}, the first term in \eqref{NewHamiltonian} becomes
\begin{equation}
    \frac{1}{4K^{(2)}\sum_y\phi_y^k\sum_z\phi_z^k}\omega^i_x\omega^i_x \,.
    \label{KineticTerm}
\end{equation}
We define $1/\sum_y\phi^k_{y\neq x}\sum_{z\neq x}\phi^k_z=N_x$. Since $\phi^i_x\propto \chi^i_x$, we have the freedom to choose the ratio $l$, the square of which we can set to be $l^2=N_x$, then defining $N_x/K^{(2)}$ as the lapse function. The temporal Hamiltonian constraint then becomes 
\begin{equation}
    \frac{N_x}{4K^{(2)}}\biggl(-\omega^i_x\omega^i_x+(K^{(2)})^2\sum_{y,z}\omega^i_x\omega^i_y\phi_y^j\phi^j_z\omega_z^k\omega^k_x\biggr)\,.
    \label{LeeCons1}
\end{equation}
For the remaining term, using \eqref{relation2} we find
\begin{equation}
    \frac{K^{(1)}}{2K^{(2)}}\varphi(k_x)\pi(k_x)=\frac{K^{(1)}}{2K^{(2)}}\phi^i_x\omega^i_x\,, \label{LeeCons2}
\end{equation}
which is nothing but the spatial Hamiltonian constraint of Ref.~\cite{Lee:2019uix}. 
Combining Eq.~\eqref{LeeCons1} and Eq.~\eqref{LeeCons2}, we have been then able to reproduce the Hamiltonian of the emergent gravity theory proposed in \cite{Lee:2019uix}. If the reflection symmetry is recovered, $K^{(1)}$ is switched off, and we observe that the term \eqref{NewHamiltonian} reduces to \eqref{MultiScalarH}.

\section{Entanglement state and its evolution}\label{SecEvo}

In the previous section we have introduced the Hamiltonian for a single vertex. Here we delve into the Hamiltonian operator, whose eigenstates are the GFT EG-states. To these latter states, one then assigns $\lambda_{xy}$ encoding the entanglement between vertices. A GFT functional mode defined at a vertex $x$ evolves in $\lambda_x$, which then works as a clock; the evolution of the GFT states, at each vertex, is provided by the Schr\"odinger equation \cite{Wilson-Ewing:2018mrp}. Since each vertex evolves in the same way, one expects to evolve the entire GFT EG-state by the same Schr\"odinger equation, that is 
\begin{equation}
    \hat{H}|\Psi_\Gamma\rangle = i\hbar\frac{d|\Psi_\Gamma\rangle}{d\lambda} \,.
    \label{schrodeq}
\end{equation}
With this purpose, we revise the Hamiltonian that is able to generate the multi-fingered evolution for the EG-state, assuming the Schr\"odinger equation \eqref{schrodeq} to hold, and then inspecting the consequences for the Hamiltonian.

The action of the projector operator, defined in Eq.~\eqref{Pop}, on a GFT state, in turn defined in Eq.~\eqref{GFTstate}, results in
\begin{align}
    &\mathbb{P}_i^{x\bigotimes y}|\psi\rangle \notag\\
    =&{\rm sym} \biggl(\int \prod_{z}d\mathbf{g}_z dh^{xy}_i d\lambda_x d\lambda_y \psi(\mathbf{g},\boldsymbol{\lambda})|g_i^x \rangle|\lambda_y\rangle \langle g_i^x h_i^{xy}|\langle \lambda_y|\notag\\
    &\bigotimes |g_i^y\rangle|\lambda_x\rangle\langle g_i^y h_i^{xy}|\langle\lambda_x|\bigotimes_z\Bigl(|\mathbf{g}_z\rangle\bigotimes|\lambda_z\rangle\Bigr)\biggr)\notag\\
    =& {\rm sym} \biggl(\int \prod_{z}d\mathbf{g}_z \psi(\cdots, g_i^x (g_i^y)^{-1},\cdots;\boldsymbol{\lambda})|\langle \lambda_x|\lambda_y\rangle|^2 \bigotimes_z\Bigl(|\mathbf{g}_z\rangle\bigotimes|\lambda_z\rangle\Bigr)\biggr) \,,\label{ProG}
\end{align}
having introduced $|\langle \lambda_x|\lambda_y\rangle|^2=|\langle \lambda_{xy}|\lambda_{xy}\rangle|^2$ as a constant multiplier in the functional at the integrand of Eq.~\eqref{GFTstate}, in addition to the convolution in the group elements. For an infinitesimal clock step, the evolution of the entanglement state can be written according to the Schr\"odinger equation as
\begin{equation}
    |\Psi_\Gamma(\lambda')\rangle=e^{-i\hat{H}\delta\lambda/\hbar}|\Psi_\Gamma;\lambda\rangle \,,
    \label{GraphstateS}
\end{equation}
where $\lambda$ labels the set of all the relational clocks, $\delta\lambda=\lambda'-\lambda$ is the running of the clock at each vertex, and the inner product between two states is expressed by
\begin{equation}
    \langle\Psi_{\Gamma'}|\Psi_\Gamma\rangle =\langle \Psi_{\Gamma'};\lambda'| e^{-i\hat{H}\delta\lambda/\hbar}|\Psi_\Gamma;\lambda\rangle\,. \label{collapse}
\end{equation}
We first consider the initial state to be a bare state with no links. The Hamiltonian operator not only does generate the change in the relational clock, but also creates links. By analogy to the Lee's model of gravity and to the Hamiltonian formulation of GR, we split the Hamiltonian of the system into two parts related to spacial and temporal diffeomorphisms, and into a mass term, proposing the ansatz 
\begin{equation}\label{FullHamil}
    H = N_l^{xy}\biggl(-\frac{1}{4K^{(2)}}\pi_x\pi_y+\frac{K^{(2)}}{4}k_x\phi_xk_y\phi_y\biggr)+N_s^{xy}\biggl(\frac{K^{(1)}}{2K^{(2)}}\pi_x\phi_y\biggr)+\frac{K^{(2)}}{4}M^2\phi_x\phi^x \,,
\end{equation}
where $N_l^{xy}$ is the lapse tensor determining the running of the clocks, and $N_s^{xy}$ is the shift tensor encoding the changing of the connectivity. The candidate for $N_s^{xy}$ is provided by the projection operator $\mathbb{P}^{x\bigotimes y}$, the action of which introduces an entry $N_s^{xy}=|\langle \lambda_x|\lambda_y\rangle|^2$. The quantity $N_l^{xy}$ is an auxiliary matrix, the entries of which are frame dependent scalars. When the lapse tensor and the shift tensor are both diagonal, the Hamiltonian \eqref{FullHamil} turns out to be the sum over all the Hamiltonians at each vertex. The Schr\"odinger equation then describes independent evolution at each vertex.

\subsection{Path-integral and renormalisation group}

In Eq.~\eqref{FullHamil} the second term in the temporal diffeomorphism and the mass term can be combined to a single term, which we denote as $\frac{m^2}{2}\phi_x\phi_y N^{xy}$, where $N^{xy}=N_l^{xy}k^xk^y+M^2\mathbb{I}$, $\mathbb{I}$ denoting the identity matrix. For an arbitrary state $|\Psi\rangle$, we may then consider 
\begin{equation}
    \langle 0 |e^{-i\hat{H}\delta \lambda}|\Psi\rangle \,,\label{PartiH}
\end{equation}
where $|0\rangle =\int \mathcal{D}\lambda|\lambda\rangle$, where $\lambda$ corresponds to the degrees of freedom $\phi$ defined in Sec.~\eqref{LeeEmergent}. If we define $\phi=\sqrt{N^{xy}\phi_x\phi_y}$ and combine the mass term with the state $|0\rangle$, the reference action becomes $|0'\rangle = \int \mathcal{D}\lambda e^{\frac{m^2}{2}\phi^2(\lambda)}|\lambda\rangle$, which is the same as in Eq.~\eqref{PartiexamQRG}. We can further consider $|\Psi\rangle =\int \mathcal{D}\lambda e^{S[\phi(\lambda)]}|\lambda\rangle$, which can always be achieved by defining the GFT functional $\Psi$ in Eq.~\eqref{GFTstate} as an exponential functional. We then see that Eq.~\eqref{PartiH} has the same formula as Eq.~\eqref{PartiexamQRG}, with the same action as \eqref{AcexQRG}. As a result, $e^{-i\hat{H}\delta \lambda}|\Psi\rangle$ provides the coarse-graining of the action $S[\phi]$, which means the GFT functional $|\Psi\rangle =\int \mathcal{D}\lambda e^{S[\phi(\lambda)]}|\lambda\rangle$ is deformed. Such a coarse-graining provides the quantum RG introduced in Sec.~\eqref{SecQRG}. The state $|0\rangle$ together with the mass term provides an IR fixed point, as it does not evolve according to the Hamiltonian. 
Instead, the coarse-graining is directly introduced by the Schr\"odinger equation, applied step by step in the degrees of freedom encoded by $\lambda$. Given that the Schr\"odinger equation generates a RG flow, we may delve into the path-integral formulation and the partition function of our theory.

The physical inner product between two states provides a path-integral formulation of the dynamics. Since $\lambda$ is constructed out of local DOF $\chi^i$, the path evolving a state from $\lambda$ to $\lambda+\delta\lambda$ has many possible choices. To distinguish among the different paths, we use the variables $\lambda_{xy}=\chi^i_x\chi^i_y$, which also encodes the connectivity, instead of $\lambda_x=\sum_y\lambda^i_x\lambda_y^i$.

In Eq.~(\ref{GraphstateS}), we have decomposed the state into a time-dependent phase and a stationary state component. The stationary state can be expressed by \cite{Colafranceschi:2020ern}
\begin{equation}
    |\psi_\Gamma\rangle = \sum_{J} \psi_J(\lambda_{xy}) |\theta_{\Gamma,J}(\{g_{xy}\});\lambda_{xy}\rangle\,, \label{GSexpand}
\end{equation}
where $\theta_{\Gamma,J}(\{g_{xy}\})$ is a collection of all the basis elements at each vertex $\theta^{\{j_i^{xt_i(x)}\}\vec{\iota}}_{\Gamma\{n_i^x\}{\rm open}}(\{g_l\})$, and 
\begin{equation}
    \theta^{\{j_i^{xt_i(x)}\}\vec{\iota}}_{\Gamma\{n_i^x\}{\rm open}}(\{g_l\})=\mathcal{I}_{\vec{\mathbf{\rm m}}}^{\{j_i^{xt_i(x)}\}\vec{\iota}}\times \prod_{x,i:t_i(x)=0}\sqrt{d_{j^x_i}}D^{j_i^x}_{m_i^x n_i^x}(g_i^x)\prod_{i,x}\sqrt{d_{j_i^{xt_i(x)}}}D_{m_i^x m_i^{t_i(x)}}^{j_i^{xt_i(x)}}(g_i^x (g_i^{t_i(x)})^{-1}) 
    \label{cee}
\end{equation}
are the basis elements that span the Hilbert space recovered by the Peter-Weyl transform. In Eq.~\eqref{cee}, $i$ labels the color of the link, $x$ labels the vertex, $t_i(x)$ is the target of the link incident from $x$, $j^{xt_i(x)}_i$ is the group representation on the link connected by $x$ and $t_i(x)$, `open' denotes unconnected open links, $n_i, m_i$ are the magnetic numbers defined on the open links, $\mathcal{I}$ represents the intertwiner, and $D$ is the Wigner matrix. The scalar product between the basis states $|\theta_{\Gamma,J}\rangle$ and $|\theta_{\Gamma',J'}\rangle$, with different combinatorial structure provided by $\Gamma$ and $\Gamma'$, respectively, is shown in \cite{Colafranceschi:2020ern} to correspond to the expression
\begin{equation}
\begin{aligned}
    \langle \theta_{\Gamma',J'}|\theta_{\Gamma,J}\rangle=\prod_x\delta_{j_i^{'xt_i'(x)},j_i^{xt_i(x)}}\prod_{x:t_i(x)=0,t_i'(x)\neq 0}\delta_{n_i^{t'_i(x)},n_i}\\
    \times \prod_{x:t_i(x)\neq 0,t'_i(x)=0} \delta_{n_i^{'x},n_i^{'t_i(x)}}\prod_{x:t_i(x)=t_i'(x)=0}\delta_{n_i^{'x},n_i^x}\delta(\vec{\iota},\vec{\iota}')\,.\label{GraphProduct}
\end{aligned}  
\end{equation}
This means that graph states with different combinatorial structures are not necessarily orthogonal to one another, thus graph-changes can be encountered over the evolution. Since $\lambda\in \mathbb{R}$ is a scalar coupled to the functional on the vertices, and the base vectors $|\theta_{\Gamma,J}(\{g_{xy}\})\rangle$ are derived only from the contribution that arises from the group elements $g_{xy}$, the base vectors $|\lambda_{xy}\rangle$ in Eq.~\eqref{GSexpand} are independent from the basis elements $|\theta_{\Gamma,J}(\{g_{xy}\})\rangle$. Being the two Hilbert spaces independent, we can write $|\theta_{\Gamma,J}(\{g_{xy}\});\lambda_{xy}\rangle=|\theta_{\Gamma,J}(\{g_{xy}\})\rangle \bigotimes |\lambda_{xy}\rangle$, with the value of $\lambda_{xy}$ providing the strength of the entanglement between the vertices $x$ and $y$. The wave-function $\psi_\Gamma(\lambda_{xy})$ is a functional of all the variables assigned to the links, and the states can be written in the momentum space as
\begin{equation}
    |\Psi_\Gamma\rangle = \int \mathcal{D}ke^{-ik^{xy}\lambda_{xy}}\psi(k_{xy})\prod_{A_{xy}=1}|\theta_\Gamma(\{g_{xy}\});k_{xy}\rangle\label{StateM}\,.
\end{equation}

As an analogue to spin foam, the inner product between the two graph states can be regarded as a sum over quantum histories, a series of virtual time-slices being inserted among the boundaries. For states supported on any pair of graphs $\Gamma$ and $\Gamma'$, the physical scalar product of the theory, implementing on the bulk fluctuations supported on different graphs --- i.e.~virtual changes of topology --- becomes
\begin{equation}
    \langle\Psi_{\Gamma'}|\Psi_\Gamma\rangle=\int \mathcal{D}k \mathcal{D}\theta\langle\Psi_\Gamma'|e^{-i\hat{H}\delta\lambda}|\theta^{(N-1)};k^{(N-1))}\rangle    \cdots\langle \theta^{(1)};k^{(1)}|e^{-i\hat{H}\delta\lambda}|\Psi_\Gamma \rangle \,, 
    \label{pak}
\end{equation}
where $k^{(n)}$ are the basis elements for the expansion of the interpolating states on the $n$-th slice --- $k^{(0)}$ correspond to the basis elements necessary to expand the boundary state $\Psi_\Gamma$, $k^{(N)}$ are the basis elements for the expansion of $\Psi_{\Gamma'}$, and the functional measure reads $\mathcal{D}k=\prod_{n,x,y}k_{xy}^{(n)}$ --- a similar definition will be provided below for $\mathcal{D}\lambda$. The measure $\mathcal{D}\theta$ in the path-integral denotes an integral over all the possible virtual combinatorial structures, at each slice, and $\theta^{(n)}$ corresponds to $\theta_{\Gamma,J}$ at the $n$-th slice, hence specifying the combinatorial structures. The integral over the combinatorial structures can be replaced by an integral over the shift tensor, which we have defined as encoding the shift of the graph, and denoted as $N_s^{xy}$. The integral over the shift tensor means all the possible choices of the combinatorial structures that are allowed at each step of the evolution are summed over. Integrating out the lapse tensor enables us to include all the possible choices of the clock running at each step. As a result, the basis elements $|\theta^{(n)}\rangle$ for the virtual states can be removed in Eq.~\eqref{pak}, the path running over different combinatorial structures in the path-integral and being constrained by the integral over the lapse tensor and the shift tensor. 

A specific selection of basis elements $k$ provides a possible path in the variables' configuration space. Thus, Eq.~\eqref{pak} instantiates a path-integral to evaluate which we insert a series of $\int \mathcal{D}\lambda|\lambda^{(n)}\rangle\langle\lambda^{(n)}|$ that finally provides the expression:
\begin{equation}
\begin{aligned}
    \langle \Psi_{\Gamma'}|\Psi_\Gamma \rangle =&\int \mathcal{D}k\mathcal{D}\lambda \mathcal{D}N_l\mathcal{D}N_s\langle\Psi_{\Gamma'}|k^{(N-1)}\rangle \langle k^{(N-1)}|e^{-i\hat{H}^{(N-1)}\delta\lambda}|\lambda^{(N-1)}\rangle\\
    &\times \langle\lambda^{(N-1)}|k^{(N-2)}\rangle \cdots \langle k^{(1)}|e^{-i\hat{H}^{(1)}\delta\lambda}|\lambda^{(1)}\rangle \langle \lambda^{(1)}|\Psi_\Gamma \rangle \,. \label{abo}
\end{aligned}
\end{equation}
The kernels in Eq.~\eqref{abo} are \[\langle\lambda^{(n)}|k^{(n-1)}\rangle=e^{ik^{(n-1),xy}\lambda_{xy}^{(n)}},\] and \[\langle k^{(n)}|e^{-i\hat{H}^{(n)}\delta\lambda}|\lambda^{(n)}\rangle=e^{-i\sum_x H(\lambda^{(n)}_x)\delta\lambda}e^{-ik^{(n),xy}\lambda_{xy}^{(n)}},\] hence
\begin{equation}
\begin{aligned}
    \langle k^{(n)}|e^{-i\hat{H}^{(n)}\delta\lambda}|k^{(n-1)}\rangle&=e^{-i\sum_x H(\lambda^{(n)}_x)\delta\lambda-\sum_{xy}\lambda_{xy}^{(n)}(k_{xy}^{(n)}-k_{xy}^{(n-1)})}\\
    &=e^{-i\sum_x H(\lambda^{(n)}_x)\delta\lambda-\sum_{xy}\lambda_{xy}^{(n)}\p_\lambda k_{xy}^{(n)}\delta\lambda}\label{Kernelcal}\,\\
    &=e^{-i\sum_{x,y,n}(\lambda_{xy}^{(n)}\p_\lambda k_{xy}^{(n)}+H(\lambda_{xy}^{(n)},k_{xy}^{(n)}))\delta\lambda}\,,
\end{aligned}
\end{equation}
which provides the result for the inner product (\ref{collapse}), i.e.
\begin{equation}
\langle\Psi_{\Gamma'}|\Psi_\Gamma\rangle=\int\mathcal{D}\lambda\mathcal{D}k\mathcal{D}N_l\mathcal{D}N_s\psi_\Gamma(k_{xy})\psi_{\Gamma'}(k'_{xy})e^{iS[N_l, N_s]} \,\,, \label{P-I}
\end{equation}
where the action is expressed as
\begin{equation}
    S[N_l, N_s]=\int d\lambda (-\sum_{xy}k_{xy}\p_\lambda\lambda_{xy}+H(N_l^{xy},N_s^{xy},\lambda_{xy},k_{xy}))\label{Finalaction}\,.
\end{equation}
 As a result, defining the temporal constraint as $C_0$ and the spacial constraint as $C$, respectively, the Hamiltonian is the sum of the two quantities
\begin{equation}
\begin{aligned}
    &{\rm tr}\{N_lC_0\}=N_l^{xy}\biggl(-\pi_x\pi_y + (K^{(2)})^2k_x\phi_xk_y\phi_y\biggr)\,,\\
    &{\rm tr}\{N_sC\}=N^{xy}_s\pi_x\phi_y\,,
    \label{tracedCons}
\end{aligned}
\end{equation}
and thus the action becomes
\begin{equation}
    S[N_s, N_l]=\int -d\lambda \left( \sum_{xy}k_{xy}\p_\lambda\lambda_{xy}-{\rm tr}\{CN_s\}-{\rm tr}\{C_0N_l\}\right) \,.
    \label{Sconstraints}
\end{equation}
We emphasize that this is the action \eqref{LeeAction} derived in the emergent gravity model proposed in \cite{Lee:2019uix}. Here, $k_{xy}$ play the role of collective variables, encoding the structure of entanglement. As discussed in the Lee's emergent gravity model \cite{Lee:2019uix}, the collective variables determine the dimension, the topology and the geometry of the system. Collective variables are also used in calculating mutual information between sites \cite{Lee:2016uqc}, a measure of the entanglement encoded by the relational clock. Since the stronger (weaker) mutual information between two vertices implies shorter (longer) distances, the variables $\lambda_{xy}$, whose conjugate momenta are the collective variables, are related to the lengths among the vertices.

The fact that the Hamiltonian is the generator of the RG flow was pointed out also in the Lee's model \cite{Lee:2019uix}, which we have already reproduced in the previous sections. Consequently, we can claim that the Schr\"odinger equation \eqref{schrodeq} is the RG equation, consistent with the fact that this can always be treated as a heat equation.

As shown in section~\ref{GRpart}, we can associate the fields $\phi^i_a$ to the Ashtekar connection $E^i_a$, in turn proportional to the tetrads $e^i_a$. Within the state $|\Psi\rangle$, the mode at each vertex is $\varphi_x=g_{ij}\phi^i_x\sum_y\phi^j_y$, where $g_{ij}$ denotes the metric for the internal indices. Thus $\varphi_x\propto g_{ij}e^i_a e^j_b=g_{ab}$ is the metric of the physical spacetime, and the Shr\"odinger equation then describes the flow of the metric. The heat flow equation becomes a simplified Ricci flow. The RG flow enables to deploy quantum mechanics in the description of gravity, its infrared limit providing GR as theory at equilibrium. In \cite{Lee:2019uix}, Lee showed that the state $|0\rangle$ always exists as an IR fixed point --- we obtained a similar result at the beginning of this section. The existence of the UV fixed point is still the matter of future investigations.

In Eq.~\eqref{Hamiltonian1}, the mass term corresponds to 
\begin{equation}
    \biggl(-\frac{K^{(0)}}{K^{(2)}}-\frac{(K^{({1})})^2}{2(K^{(2)})^2}\biggr)\varphi_x^2 \equiv M^2\varphi^2\,, \label{corr}
\end{equation}
which can be regarded as a sum over all the mass terms of each coupling DOFs for all the pairs of sites. We emphasize that the scale of the mass term is independent from the choice of the kernel. Setting $\delta\lambda\sim l_p$, the correspondence in Eq.~\eqref{corr} still holds, and the emergent mass term turns out to be in the vicinity of the Planck scale. In \cite{Wilson-Ewing:2018mrp,Gielen:2020fgi}, it was pointed out that, to derive the correct classical limit in cosmology, a Planckian value for the scale of the mass must be picked out. Nonetheless, in our case the scale of the mass term can be any, and its value is recovered in full generality according to the RG flow generated by the Schr\"odinger equation.

\section{Relation to the Hamiltonian formalism of general relativity}\label{GRpart}

Foliating the spacetime in a series of time-slices, the Einstein-Hilbert action can be recast as \cite{PhysRev.116.1322}
\begin{equation}
    S=\int dtdx^3\bigl(\pi^{ab}\dot{q}_{ab}-NC(\pi,q)-2N^aC_a(\pi,q)\bigr) \,,
    \label{HamilGR}
\end{equation}
where $q_{ab}$ is the induced metric, $\pi^{ab}$ is the conjugate momentum of $q_{ab}$, $N, N^a$ are the Lagrangian multipliers called the lapse function and the shift vector, respectively. The three quantities $q_{ab}$, $N$ and $N^a$ are defined by the ADM form of the metric
\begin{equation}
    ds^2=-(N^2-N_aN^a)dt^2+2N_adx^adt+q_{ab}dx^adx^b \,.
\end{equation}
Above, $C, C_a$ are the time constraint and the spatial constraint, respectively, defined by
\begin{align}
    &C=G_{abcd}\pi^{ab}\pi^{cd}-\sqrt{q}R[q]\,,\\
    &C_a=D_a\pi^{ab} \,,
\end{align}
where $G_{abcd}=\frac{1}{2\sqrt{q}}(g_{ac}g_{bd}+g_{ad}g_{bc}-g_{ab}g_{cd})$ is the DeWitt super-metric, and $D_a$ denotes the covariant derivative.

Using the Ashtekar variables \cite{PhysRevLett.57.2244}
\begin{equation}
    A^i_a=\Gamma^i_a[e]+\beta k^i_a \,,
    \label{AshtekarV}
\end{equation}
where $\Gamma^i_a[e]$ is the torsion-less spin connection defined by $\Gamma^i_a[e]={\epsilon^i}_{jk}\Gamma^{jk}_a$, with $\Gamma^{ij}_a$ being Christoffel symbols, $k^i_a$ is the extrinsic curvature, and $\beta$ is the Immirzi parameter. If $\beta$ is purely imaginary, $A^i_a$ provides a complex Ashtekar connection, with a considerable exemplification of the form of the Hamiltonian constraints. For real values, $\beta$ has been refereed to in the recent literature as the Barbero-Immirzi parameter $\gamma$, entailing a more complicated form of the Hamiltonian constraints. As an illustrative example, we will assume $\beta=i$ while showing that the Hamiltonian we are considering is equivalent to the Ashtekar's Hamiltonian. 

The conjugate momentum of $A^i_a$ is the densitized triad,
\begin{equation}
    E^i_a=|{\rm det}(e)|e^i_a\,.
    \label{AshtekarM}
\end{equation}
In the light of this change of variables, the Hamiltonian constraints can be written as 
\begin{align}
    &H[N]=\int dx^3 N \epsilon_{ijk}F^i_{ab}E^{aj}E^{bk} \,, \label{AshtimeC} \\
    &D[\vec{N}]=\int dx^3 N^a F_{ab}^iE^{bi}\,, \label{AshspaC}
\end{align}
where $F^i_{ab}=\p_aA^i_b-\p_b A^i_a+\epsilon^i_{jk}A^j_aA^k_b$ is the curvature of the Ashtekar connection. An additional constraint generating the gauge transformations, the Gauss constraint, is introduced by using the Ashtekar variables, i.e.   
\begin{equation}
    G[\Lambda]=\int dx^3 \Lambda^i D_a E^a_i \,.
    \label{AshGaussC}
\end{equation}
Our action (\ref{Finalaction}) has a similar structure to the GR action (\ref{HamilGR}): we can recast the former one in terms of the Ashtekar variables, as a combination of \eqref{AshtimeC}-\eqref{AshGaussC}.

We constrain our model in 3+1 dimension, where the $3-d$ spacial variables are provided by the coupling DOF, and the temporal component is a parameter that selects the time-slice on which the variables are defined. We intend to promote this parameter to be our relational clock, corresponding to the parameter deployed to represent the total entanglement $\lambda$. From now on, we use the labels $a,b,c$ for the spacial indices, and $i,j,k$ for the internal indices. The variables $\phi^i_x$ are the one-form valued internal vectors at the point $x$, i.e. $\phi^i_x=\phi^i_{x,a}dx^a$, and similarly, $\omega^i_x=\omega^i_{x,a}dx^a$. We can assign a Lie group structure to the internal labels --- the label $i$ denotes a component of the group generator $T^i$ with Lie bracket $[T^i, T^j]={f^{ij}}_kT^k$. The terms within \eqref{Hamiltonian1} can be recast considering the background topology as a $G$-bundle, and selecting the structure of the Lie group to be $\SU(2)$. For the second term in the Hamiltonian \eqref{Hamiltonian1}, from \eqref{KineticTerm} and using \eqref{momentum}, we find 
\begin{equation}
\begin{aligned}
    \pi_x^2&=\omega^i_x\phi_{x,i}\omega^j_x\phi_{x,j}/\varphi_x^2\\
    &={\rm tr}\{[\omega^i_x\wedge\omega^j_x]\cdot [\phi_{x,i}\wedge\phi_{x,j}]\}/\varphi_x^2\\
    &={\rm tr}\{{\epsilon^{ij}}_kT^k\omega^i_{x,a}\omega^j_{x,b}\epsilon_{ijk}T^k\phi_{x,i}^a\phi^b_{x,j}\}/\varphi^2_x \,, \label{Kash}
\end{aligned}
\end{equation}
where ``${\rm tr}$'' denotes the trace over the group generators $T^k$, and we have changed the direct products of two $\omega$ and two $\phi$ to the wedge product for the Lie group valued 1-forms, and the unit 1-forms $dx^a$ are canceled by the inner product. Recall that in the derivation of our model, $\omega$ plays the role of a momentum, and $\omega^i$ are proportional to $\p^i$. It is therefore straightforward to deduce that $\omega_a$ is proportional to $\p_a$, and that on the principal fiber bundle it becomes the exterior covariant derivative $D_a$. Consequently, we find $\epsilon^{ijk}\omega^i_a\omega^j_b=\epsilon^{ijk}\omega^iD_a\omega^j_b=F_{ab}^k$ to be the gauge strength for $\omega$. The time constraint then becomes
\begin{equation}
    H[N]=\int dr^3 \biggl(\frac{1}{2K^{(2)}\varphi^2(\mathbf{r})}\biggr)\epsilon_{ijk}F_{ab}^k(\mathbf{r})\phi^{ia}(\mathbf{r})\phi^{ib}(\mathbf{r}) \,. \label{Hu}
\end{equation}
We have substituted the sum over the sites $x$ with the integration over the space in the continuum limit. If we define the lapse function as $N(\mathbf{r})=1/2K^{(2)}\varphi^2(\mathbf{r})$, Eq.~\eqref{Hu} acquires the same form as \eqref{AshtimeC}. This expression of the lapse function is generally position dependent: time changes at different space-time points are different.

To properly deal with the term that is proportional to $\pi_x\phi_x$ and plays the role of the spacial constraint, we exploit the shift tensor $N_s^{xy}$. The spacial constraint then becomes proportional to $N^{xy}_s\phi_x\pi_y$. Expanding $\pi_y$ around $\pi_x$, we finally recover 
\begin{equation}
\begin{aligned}
    C_s&={\rm tr}\{N^{xy}\phi_x^i(\omega_{x,i}+\p_a\omega_{y,i}|_{y=x}(r^a_y-r^a_x))\}\\
    &=\sum_{x,y}(N^{xy}\phi_x^i\omega_{x,i}+N^{xy}\phi_x^i\p_a\omega_{y,i}|_{y=x}\delta r^a_{xy})\\
    &=\sum_x N_x\phi_x^i\omega_{x,i}+N_x^a\phi_x^iD_a\omega_{x,i}\,.\label{SCons}
\end{aligned}
\end{equation}
In the last line, we have defined $N_x=\sum_y N_{x,y}$ and $N^a_x=\sum_y N_{xy}\delta r^a_{xy}$, with $\delta r^a_{xy}=(r^i_y-r^i_x)$, and replaced the ordinary derivative with the covariant derivative, corresponding to the parallel transport on the bundle. The second term in the last line of Eq.~\eqref{SCons} becomes
\begin{equation}
    D[\vec{N}]=\int dr^3 N^a(\mathbf{r})\phi^b_i(\mathbf{r}) F_{ab}^i(\mathbf{r})\,,
\end{equation}
with the same form of the spatial Hamiltonian constraint in Eq.~\eqref{AshspaC}. The first term in Eq.~\eqref{SCons} is expressed by
\begin{equation}
\begin{aligned}
    G[\Lambda]&=\int dr^3 N(\mathbf{r})\omega_i(\mathbf{r}) D_a\phi^{a,i}(\mathbf{r})\\
    &=\int dr^3 \Lambda_i(\mathbf{r}) D_a\phi^{a,i}(\mathbf{r}) \,,
\end{aligned}
\end{equation}
where we have defined $ \Lambda_i=N(\mathbf{r})\omega_i(\mathbf{r})$, and hence reproduces the form of the Gauss constraint in Eq.~\eqref{AshGaussC}. 

Comparing the new Hamiltonian and Gauss constraints to the ones in the Ashtekar's formalism, we can deduce that $\phi^i_a$ plays the role of the densitized triad $E^i_a$, and $\omega^i_a$ provides the Ashtekar connection. Since $\omega^i_a$ given by eq.\eqref{momentum} is the conjugate momentum of $\phi^i_a$, it is easy to check that the Poisson bracket $\{\phi^i_{a,x},\omega_j^{b,y}\}=\delta^i_j\delta^b_a\delta^y_x$ is closed. Nonetheless, the Hamiltonian constraints is equivalent to Ashtekar's one only when the Immirzi parameter is pure imaginary. For a real Barbero-Immirzi parameter, the extra term proportional to $(1+\gamma^2)$ vanishes, as discussed in Ref.~\cite{Alexander:2012ge} in relation to the chiral symmetry breaking. The internal symmetry of the coupled DOF in our model is ${\rm O}(N)$, where $N$ is a large number for local DOF. The $\SU(2)$ internal symmetry should then originate from a symmetry breaking of ${\rm O}(N)$, as discussed by Lee in Ref.~\cite{Lee:2019uix}. Providing a full Ashtekar's Hamiltonian for any value of the Barbero-Immirzi parameter is beyond the scope of this work, and we will abstain from that. The fact that the form of the Hamiltonian constraints (within the Hamiltonian formalism of GR) is reproduced does not ensure these to act like symmetry constraints in GR. The vanishing of the symmetry constraints and the closure (on-shell) of their algebra is rather attained at the equilibrium, i.e.~in the infrared limit of the RG flow. 

The existence of the infrared limit of the RG flow ensures that this model of emergent gravity, originating from entangled local DOF, provides at equilibrium the Einstein's theory of gravity. On the other hand, the bridge between the theory of entangled local DOF formulated by Lee and Einstein's GR, expressed in the Ashtekar variables and with coupling constant determined by the mass term in Lee's theory, is realized by GFT. The existence of a UV fixed point is still under investigation. A recent attempt toward this direction is provided by \cite{lulli2021stochastic}, in which the Ricci flow generates a geometric RG flow, its physical meaning being related to relaxation processes in out-of-equilibrium quantum field systems. 

We remark that in order to deal with the spacial constraint we have used the shift tensor. Nonetheless, we did not use the lapse tensor to constrain the trajectories in the time-reparametrization constraint. This is consistent, since for the time-reparametrization constraint the lapse tensor is diagonal: the choice of the frame, i.e.~the way to decompose the Hilbert space in the Lee's theory enables the vertices to evolve independently into one another.

\section{Quantization of the Hamiltonian}

The Hamiltonian \eqref{NewHamiltonian} can be written as
\begin{equation}
\hat{H}=\frac{K^{(2)}}{2}\biggl(\frac{\hat{\pi}^2}{(K^{(2)})^2}+\varpi^2(k)\hat{\varphi}^2\biggr)+i\frac{K^{(1)}}{K^{(2)}}\hat{\varphi}\hat{\pi} \,,\label{HamilOperator}
\end{equation}
where $\varpi^2(k)=m^2+k^2$, and $m^2$ is defined by the prefactor of the last term in \eqref{NewHamiltonian}. We can quantize the fields $\varphi$ and $\pi$ by defining the ladder operators \cite{Gielen:2020fgi}
\begin{align}
    &\hat{a}_J(k)=A_J(k)\varphi_J(k)+\frac{i}{2A_J}\hat{\pi}_J(k)\label{annihilationO}\,,\\
    &\hat{a}^\dagger_J(k)=A_J(k)\varphi_J(k)-\frac{i}{2A_J}\hat{\pi}_J(k)\label{creationO}\,,
\end{align}
where $A_J^2(k)=|\varpi(k)| |K^{(2)}|/2$. Substituting these definitions into the Hamiltonian operator \eqref{HamilOperator}, we have
\begin{equation}
    \hat{H}=\sum_J\biggl({\rm sgn}(K_J^{(2)})\frac{|\varpi_J(k)|}{4}(\hat{a}_J^\dagger\hat{a}_J+\hat{a}_J\hat{a}_J^\dagger)-\frac{K^{(1)}}{2K^{(2)}}((\hat{a}_J^\dagger)^2-\hat{a}_J^2-\hat{a}^\dagger_J\hat{a}_J+\hat{a}_J\hat{a}^\dagger_J)\biggr)\,.
    \label{HamilOp1}
\end{equation}
The ladder operators and the Hamiltonian, written in the Shr$\ddot{\rm o}$dinger picture, do not evolve with respect to the clock field. The ladder operators are subjected to the commutation relation
\begin{equation}
    [\hat{a}_J, \hat{a}_{J'}^\dagger]=\delta_{J,J'} \,,
\end{equation}
which provides for the Hamiltonian operator \eqref{HamilOp1} the expression
\begin{equation}
    \hat{H}=\sum_J\biggl({\rm sgn}(K_J^{(2)})\frac{|\varpi_J(k)|}{2}(\hat{a}_J^\dagger\hat{a}_J+\frac{1}{2})-\frac{K^{(1)}}{2K^{(2)}}((\hat{a}_J^\dagger)^2-\hat{a}_J^2+1)\biggr)\,. \label{Hami}
\end{equation}
The Hamiltonian operator in Eq.~\eqref{HamilOp1} is not hermitian: its eigenstates have imaginary components --- the operator $\hat{\pi}$ includes component $i\hat{\varphi}$ --- and the corresponding modes are unstable. On the other hand, the imaginary part of the Hamiltonian entails an imaginary part of the action. Even though in the path-integral formulation of quantum theories the action is customarily considered to be real, recently complex actions have been considered in the literature --- see e.g. Refs.~\cite{Nielsen:2006pz,nagao2015reality,nagao2017complex,Nagao:2017ztx,nagao2022reality}.

In \cite{nagao2017complex} it was shown that a complex action implies a future-included theory, which means that not only the past state is fixed by the action, but also the future state. For a complex Hamiltonian, the eigenstates are in general not orthogonal, i.e. the states $|\lambda_i\rangle$ satisfy $\hat{H}|\lambda_i\rangle=\lambda_i|
\lambda_i\rangle$, with $\langle \lambda_i|\lambda_j\rangle\neq \delta_{ij}$. To fix this issue, a proper inner product has been introduced in Refs.~\cite{nagao2011automatic,nagao2011formulation} that can render the eigenstates of the Hamiltonian orthogonal by means of a hermitian operator $Q$ --- $Q$ is provided by $(P^\dagger)^{-1}P$, where $P$ an operator that diagonalizes the Hamiltonian --- whose action on generic states $A$ and $B$ is defined by $I_Q=\langle A|\hat{Q}|B\rangle=\langle A|_{Q}B\rangle$. In future-included theories \cite{Nielsen:2006pz,nagao2013theory,nagao2012correspondence} the time-evolution of future states $|B(t_B)\rangle$ and past states $|A(t_A)\rangle$ is provided by the Schr\"odinger equation --- with Hamiltonian given by Eq.~\eqref{Hami} --- the observables being calculated according to 
\begin{equation}
    \langle\hat{O}\rangle^{BA}_{Q}=\frac{\langle B(t_B)|_{Q}\hat{O}|A(t_A)\rangle}{\langle B(t_B)|_{Q}A(t_A)\rangle}\,.
\end{equation}

For a complex Hamiltonian operator, the Schr\"odinger equation for future-states will be driven by the so-called $Q$-Hermitian Hamiltonian \cite{nagao2015reality} --- i.e. a Hamiltonian satisfying $\hat{H}=\hat{H}^{\dagger^Q}$ --- where $Q$-Hermitian conjugation is defined by the relation 
\begin{equation}
\hat{H}^{\dagger^Q}=\hat{Q}^{-1}\hat{H}\hat{Q}\,.
\end{equation}
In this scenario, a complex Hamiltonian provides an imaginary part for its eigenvalues, and only those eigenvalues whose imaginary parts are maximal (among all the other imaginary components) will contribute maximally to the scalar product. Past and future states are chosen such that the inner product equipped with the operator $\hat{Q}$ is maximized, according to the maximal principle in the future-included Complex Action Theory (CAT). In the periodic CAT introduced in \cite{nagao2022reality}, characterized by the boundary condition $|B(t_B)\rangle=|B(t_A+t_p)\rangle=|A(t_A)\rangle$, the maximization principle will provide some typical values for the period $t_p$ that maximize the scalar product.

Novel insights for CAT might arise from our Hamiltonian \eqref{HamilOp1}. In the path-integral \eqref{P-I}, initial and final graph states correspond to past and future states, hence are selected by the maximal principle. The change of the relational clock from the past $\lambda_i$ to the future $\lambda_f$ provides a period $\Delta\lambda$, which can be matched to the period $t_p$ in the periodic future-included theory. This finally provides a periodic space-time, with the clock $\lambda$ encoding the strength of the total entanglement. The mutual information or entropy in the system is monotone increasing, but the spacetime is periodic, implying a cyclic universe. A detailed discussion of periodic universe emerging from our model will be a matter of forthcoming studies.

\subsection{Volume operator}

In section~\ref{GRpart}, we have discussed the correspondence between our emergent Hamiltonian and the Hamiltonian formalism of GR in the Ashtekar variables. It is therefore tempting, having accomplished the quantization of the theory, to inspect relevant geometric observables within the two frameworks. In this section, we then discuss the correspondence between implementations of the volume operator in the two schemes.

The volume operator defined out of the Ashtekar variables reads \cite{dona2010introductory}
\begin{equation}
    V_R=\int_R d^3x \sqrt{g}=\int_R d^3x \sqrt{\left|\frac{1}{3!}\epsilon_{abc}\epsilon^{ijk}E^a_iE^b_jE^c_k\right|} \,,
    \label{VoperatorA}
\end{equation}
where $R$ is a specific spatial region. From the perspective we are following, we may define the volume operator in a way similar to \cite{Wilson-Ewing:2018mrp,Gielen:2020fgi}, i.e. 
\begin{equation}
    \hat{V}=\sum_J V_J\hat{a}_J^\dagger\hat{a}_J\label{VoperatorM}\,,
\end{equation}
where $V_J$ is the eigenvalue of the volume operator acting on a node of the spin-network. Even though we still use $J$ as the label of the eigenvalues, they only depend on the quantum numbers $\vec{j}$ and $\vec{\iota}$, as they are eigenvalues of a specific node. The label $J$ represents a specific configuration of the geometry. Thus, we only need to consider a single mode $J$ to check the correspondence between the two realizations of the volume operator, and will ignore the label $J$ in the following calculation.

Substituting the ladder operators \eqref{annihilationO} and \eqref{creationO} into \eqref{VoperatorM}, we find 
\begin{equation}
    \hat{V}=V\biggl(A^2\varphi^2+\frac{1}{4A^2}\pi^2-\frac{i}{2}\biggr)\,.
\end{equation}
From \eqref{Kash}, we may derive 
\begin{equation}   \varphi^2\pi^2=\epsilon_{ijk}\phi^a_i\phi^b_j\epsilon^{ijk}\omega^i_a\omega^j_b\equiv C\label{relationC}\,,
\end{equation}
which implies
\begin{equation}
    A^2\varphi^2+\frac{1}{4A^2}\pi^2=\biggl(\frac{A^2}{\pi^2}+\frac{1}{4A^2\varphi^2}\biggr)C \,.
\end{equation}
From the definition of $E^c_i$ we recover
\begin{equation}
    E^c_i=\frac{1}{2}\epsilon_{ijk}\epsilon^{abc}e^j_a e^k_b\,.\label{AshE}
\end{equation}
Since $\phi^a_i \sim E^a_i$, for the choice of the $\SU(2)$ internal group, the relation \eqref{relationC} can be revised into 
\begin{equation}
\begin{aligned}
    C=&\epsilon_{ijk} E^{aj}E^{bk} (\epsilon^i_{jk}\omega^j_a\omega^k_b)\frac{1}{2}\epsilon^{abc}e^j_a e^k_b e^a_j e^b_k \frac{1}{3}\epsilon_{abc}\\
    =&\epsilon_{abc}\epsilon^{ijk}E^c_iE^a_jE^b_k\omega_a^j\omega_b^k e^a_j e^b_k\,. 
    \label{relationC2}
\end{aligned}
\end{equation}
Some technical details must be considered to write \eqref{relationC2}. Since $\epsilon_{ijk}$ is non-vanishing only if $i,j,k$ are different from one another, and since for the $\SU(2)$ group the adjoint indices $i,j,k$ runs over $\{1,2,3\}$, the terms appearing in the \eqref{relationC} only involve the indices labelled by $\{i,j,k\}=\{1,2,3\}$. We may then introduce $E^c_i$ by using $1=\epsilon^{abc}\epsilon_{abc}\delta_j^j\delta^a_a\delta_k^k\delta^b_b/6=\epsilon^{abc}\epsilon_{abc}e^j_ae_j^ae^k_be_k^b$ and equation \eqref{AshE}. Combining all the terms, it is straightforward to check that only those terms that appear in the second line of \eqref{relationC2} will be left. Combining the previous relations, we can define 
\begin{equation}
    V=\biggl(\frac{A^2}{\pi^2}+\frac{1}{4A^2\varphi^2}\biggr)\omega^j_ae^a_j\omega^k_be^b_k\,,
\end{equation}
whose expression clarifies that the realization of the volume operator generate the same value when acting on a node.

\subsection{Cosmological implications}

The quantized Hamiltonian operator also provides insights for early cosmology. For example, the coherent state may correspond to the cosmological sector of GFT \cite{Gielen:2013kla}. The Friedman equation on flat FLRW spacetime is derived from the emergent coarse-grained dynamics of a simple coherent state in GFT, with correct classical limit \cite{Oriti:2016qtz,Oriti:2016ueo}. However, the interaction part was ignored in studies focusing on the FLRW spacetime. In our theory, the interaction is already encoded in the entanglement graph state, and the scalar product between the states is a coarse-grain process. If we consider states expanded on the coherent states' overcomplete basis, we may recover the Friedman equations in curved spacetime. By calculating the expectation value of the volume operator, the energy density could be estimated from $\epsilon=\pi^2/2V^2$. To recover the correct classical limit, the scale of the mass term should be then corresponding to the Planck mass.

Finally, we have already mentioned that the imaginary action may provide a periodic universe, within the framework of the periodic future-included CAT. Since cosmology is not the main focus of this analysis, we leave further considerations to studies in progress.

\section{Conclusions} \label{Conclu}

We have assigned multiple ultra-local DOF to each virtual vertex of a GFT graph, and defined a local scalar field including information of correlations between the DOF at different vertices, and playing the role of a relational clock. The entanglement between sites introduces causality between the vertices. By constructing an effective action from GFT, we have shown that the model of quantum gravity proposed in \cite{Lee:2019uix} emerges from the evolution of entanglement-states, and that the Fourier modes of the relational clock provide collective variables. A path-integral formulation has been provided in terms of the physical scalar product between the two GFT entanglement-graph states with different relational clocks. We have further demonstrated that the path-integral induces the RG flow generated by the Hamiltonian, that the RG equation is individuated by the Schr\"odinger equation, and that the RG scale is provided by the difference of arbitrary relational clocks. By defining and quantizing the ladder operators, we have found that the Hamiltonian is complex, and that the action introduced in the path-integral provides a future-included complex action theory (CAT), whose 'time' is provided by the entanglement strength, encoded in the relational clock. The maximal principle in CAT ensures the expectation value of operators to be real, while past states and future states are determined by the imaginary part.

For a $SU(2)$ $G$-bundle, the Hamiltonian of the theory turns out to be equivalent to the one of GR in the Ashtekar variables. This implies that the coarse-graining of the quantum entanglement has a drift-term that at equilibrium is proportional to the Hamiltonian of GR. After quantizing the Hamiltonian introducing the ladder operators, the volume operator provides the spacetime volume in GR, if a suitable eigen-volume is defined. Setting the renormalization scale at the Planck scale, the mass term in the Hamiltonian provides the Planck mass.

Detailed studies will be devoted to the cosmological implications of our model, along the lines of \cite{Wilson-Ewing:2018mrp,Gielen:2020fgi}, in which the first Friedman equation was recovered on flat FLRW spacetime. Considering the interaction between vertices, i.e. a non-vanishing  spacetime curvature, will definitely provide more interesting features for the cosmological implications. Finally, considering the future-included periodic CAT, another interesting question concerns the derivation of a cyclic universe from our model.

\vspace{0.3cm}
{\bf Acknowledgements}\\
\noindent 
The authors thank Steffen Gielen, Ling-Yan Hung, Simon Langenscheidt, Luca Marchetti, Daniele Oriti, and Edward Wilson-Ewing for valuable discussions. The work of S.A.~is supported in part by the Simons Foundation award number 896696.
A.M.~wishes to acknowledge support by the Shanghai Municipality, through the grant No.~KBH1512299, by Fudan University, through the grant No.~JJH1512105, the Natural Science Foundation of China, through the grant No.~11875113, and by the Department of Physics at Fudan University, through the grant No.~IDH1512092/001.
R.P.~is supported in part by the Swedish Research Council grant, contract number 2016-05996, as well as by the European Research Council (ERC) under the European Union's Horizon 2020 research and innovation programme (grant agreement No 668679).

\bibliographystyle{JHEP}
\bibliography{ref}

\end{document}